THE SPATIAL ACCESSIBILITY OF ATTRACTIVE PARKS IN CHICAGO AND A PROPOSED PLANNING SUPPORT SYSTEM TO EVALUATE THE ACCESSIBILITY OF POINTS OF INTEREST

BY

TIANCHEN HUANG

THESIS

Submitted in partial fulfillment of the requirements
for the degree of Master of Landscape Architecture in Landscape Architecture
in the Graduate College of the
University of Illinois at Urbana-Champaign, 2021

Urbana, Illinois

Adviser:

    Professor Brian Deal, Chair
    Visiting Assistant Professor Si Chen, University of Oklahoma
    Assistant Professor, Conor E. O'Shea

# ABSTRACT

Urban parks play an essential role in meeting the ecological, social, and recreational requirements of residents. Access to urban parks reflects people's quality of life. The present research focused on the cumulative accessibility by walking and driving to attractive urban parks for different population groups in Chicago. The present study used the ratings and the number of reviews on Google Maps to evaluate each parks' attractiveness. The results present the cumulative accessibility scores using gravity, linear, and kernel models. In addition, the spatial distribution of the accessibility to parks for population groups of different races and levels of income are shown at the 90 m X 90 m land cell scale and at the community level. Highly attractive parks that people walk to receive a high accessibility score. However, parks with high accessibility scores that people drive to were along the major highways. The present study also determined that the Black and high-income populations have a higher accessibility score to parks than other population groups. Moreover, a planning supporting system is proposed that uses rating data that can be gathered from apps such as Google Maps and Yelp to evaluate all types of Points of Interest (POIs) in parks.

# ACKNOWLEDGEMENT

Throughout the writing of the thesis, I have received a lot of support and assistance.

I would like first to thank my Thesis Chair, Professor Brian Deal. His insightful guidance in land use modeling and sustainable planning helps me to form my research topic and methodology. His valuable feedback about our routine thesis meeting pushed me to refine my thinking step by step and brought my work to a better level. Furthermore, Professor Deal was very generous and helpful in giving me plenty of suggestions and support in my future research and career.

I would like to appreciate Professor Si Chen and Professor Conor O'Shea. As my thesis committee member, Professor Chen gave me beneficial suggestions on methods and encouraged me a lot to finish my thesis. Professor O'Shea organized the thesis presentation and kept checking the progress of our thesis work in the department.

I also cherish Professor Jie Hu, Professor Stephen Sears, Professor David Hays, Professor Rolf Pendall, Professor Mary Pat McGuire, Professor Ziqi Li, Ms. Michelle Inouye, Ms. Marti Gortner, and Dr. Lori Davis. During my three years at UIUC, I worked with them, communicated with them, and took their classes. They helped me in many aspects, and the experience broadened my horizon and inspired me in study and life. It was impossible for me to finish the thesis without their support and instructions.

Finally, I would like to thank the support and help of all members from Landuse Evolution and impact Assessment Model (LEAM) laboratory and Illinois Smart Energy Design Assistance Center (SEDAC). During my one-year experience in the LEAM lab and half-year experience in SEDAC, I got to know many excellent students, researchers, and staff. I gained valuable research and work experience in the past year, and this experience contributed a lot in forming my thoughts about my thesis and the direction of future research.

# TABLE OF CONTENTS

# CHAPTER 1: INTRODUCTION

## Importance of Urban Parks

Urban parks can meet people's diverse requirements, such as for residents to rest, for entertainment, and to exercise. One of the most important elements to measure the quality of life of urban residents is their access to urban parks (Egerer et al., 2019). An urban park integrates the harmonious coexistence between humans and the ecological environment, functional space, and nature. The urban park is a recreational place for people and has ecological functions such as beautifying the city, adjusting the urban microclimate, improving the air quality, and maintaining the urban ecological balance (Brown et al., 2018). Accessibility refers to the level of difficulty involved in moving from one location to another. It reflects the spatial resistance that a population has with regards to the entire process of traveling (Hansen, 1959). As an essential factor to measure the rationality of the spatial layout of urban public service facilities, accessibility is often expressed quantitatively using the indicators distance, time, and cost (Insardi & Oliveira Lorenzo, 2019). Measuring people's accessibility to parks is, to some extent, a reflection of their living environment. Access to parks is not equal for all population groups (Dai, 2011). Determining the current spatial distribution of parks can be used as a reference for the future planning and modification of urban park systems (Dony, Delmelle, & Delmelle, 2015).

## Park Accessibility

There is much literature on measuring the attributes of parks and their accessibility. Some researchers focused on the level of service that indicates the minimum ground space needed to satisfy the actual recreation demands of residents (Oh & Jeong, 2007). It can show the service capacity of parks and people's actual requirements. Research has been completed on the classification of parks using the Analytic Hierarchy Process and analyzing the value quantitatively (Li et al., 2019). Previous studies also involved evaluating accessibility to parks, such as classifying the access to parks of different groups of the population using network analysis (Nicholls, 2001). The method used to evaluate accessibility is critical because it can identify the areas in which public services are deficient. The two-step floating catchment area method (2SFCA) is another method used to measure accessibility to parks (Hu et al., 2020). In previous research, several types of regression models were applied to calculate accessibility at the level of the neighborhood (Ma et al., 2018).

However, the methods used in previous studies were usually site-specific and time-consuming, and the utilization of limited samples restricts data collection for studies involving large areas (Chang et al., 20191101). In previous studies, the methods reflected the service capacity of the parks. However, much time was required to collect the data which involved making telephone calls to residents, and manually searching for all of the facilities in each park in one year (Mertes ,1996). Consequently, these manual methods could not be used to complete research on a metropolis with hundreds of parks. Some researchers define the value of parks by their type (Xing et al., 2018). Although the value can be calculated rapidly for hundreds of parks, it cannot reflect people's actual altitudes towards the parks. Although studies have been completed on parks based on data gathered from social media like Twitter (Hamstead et al., 2018), few of them include assessing accessibility. Previous research on measuring accessibility to parks have flaws.

Accessibility to parks can be measured using Euclidean distance or network analysis in ArcGIS (Nicholls, 2001). Euclidean distance can only show the linear distance between two locations, and does not reflect the actual distance by road, or barriers to access such as rivers, and landforms (Xingyou Zhang et al., 2011). A network analysis is better because it reflects the road network and barriers (Miyake, Maroko, Grady, Maantay, & Arno, 2010), but the calculation in ArcGIS's network analysis cannot reveal cumulative accessibility of each area being assessed. This means that when people from one location have access to multiple parks, it is treated the same as it would if they only had access to one park. Another weakness of previous research is that accessibility was usually measured in relatively large scales such as at the level of the community or neighborhood (Langford, 2013). They did not consider that residents in the same neighborhood may have different accessibilities to parks, which is especially true in large neighborhoods.

  To address the limitations of previous studies, data on the rating and the number of reviews of all parks in Chicago were obtained from Google Maps to calculate the parks' score of attraction. Gravity, linear, and kernel regression models were applied to reflect the spatial distribution and accessibility to parks in Chicago by people that are walking and driving. After being combined with census data, the accessibility of parks by different population groups was also assessed at both the 90 m X 90 m land cell scale and at the community level.

# CHAPTER 2: SITE SELECTION AND DATA COLLECTION

## Site Selection

In the present research, the City of Chicago was selected as the research area. It is the third most populous city in the United States, with an estimated population of 2,693,976 in 2019 (Insardi, 2019). Chicago is an international hub for finance, culture, commerce, industry, education, technology, telecommunications, and transportation (Gang & Wolf, 2019). Researching spatial accessibility in a metropolitan area like Chicago instead of a county or city with a smaller population and fewer parks, can assist us to better understand the applicability of the present research.

## Data Collection

In 2019, data on the parks in Chicago were obtained from the Chicago Park District. In 2019, there were a total of 583 parks in Chicago (Figure 2.1). The data included information on the location and size of each park. There are many different types of parks in Chicago which includes the many famous parks around Chicago's lakefront, such as Grant Park, and Millennium Park, and in contrast to this, some neighborhoods have small parks with very few facilities.

People can rate places on Google Maps. Data on the rating and reviews of each park were collected from Google Maps in August 2020. The rating is from 1 to 5 stars and shows people's attitudes towards the parks they visited. In 2019, there were 583 parks in Chicago, however, 43 of them did not have any reviews on Google Maps, which suggested that they were much less popular than the other parks. Consequently, information for only 540 parks could be obtained from Google Maps.

The 2019 census data for Chicago was obtained from the United States Census Bureau. The census data includes total population size, total number of households, the size of the population by race, and the income in each census tract (Figure 2.2, 2.3, and 2.4). The census data demonstrated that the population density was higher in the Central, West, and North Sides of Chicago.

In 2019, data on the road network was obtained from the United States Census Bureau. It includes the spatial distribution of the road network in Chicago and the road hierarchy. Chicago community division data were obtained from the United States Census Bureau in 2019 (Figure 2.5) and showed that Chicago is divided into nine sides and 77 communities. It shows the spatial division of Chicago at the community level. Landcover data were obtained from the U.S. Geological Survey in 2019 (Figure 2.6). All of these data were clipped in ArcGIS to remove the areas outside of Chicago.

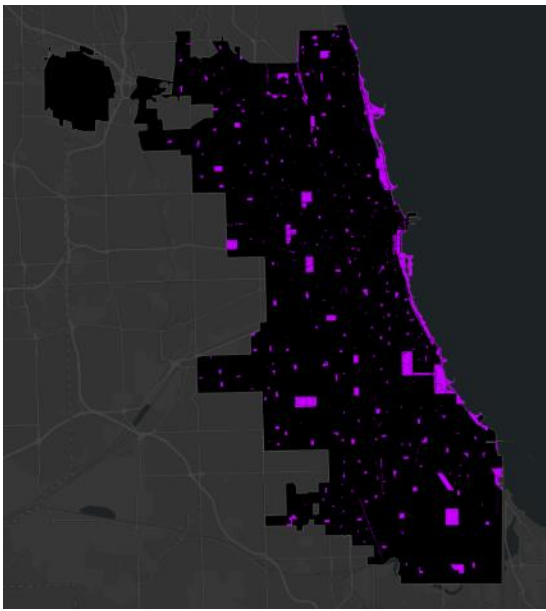 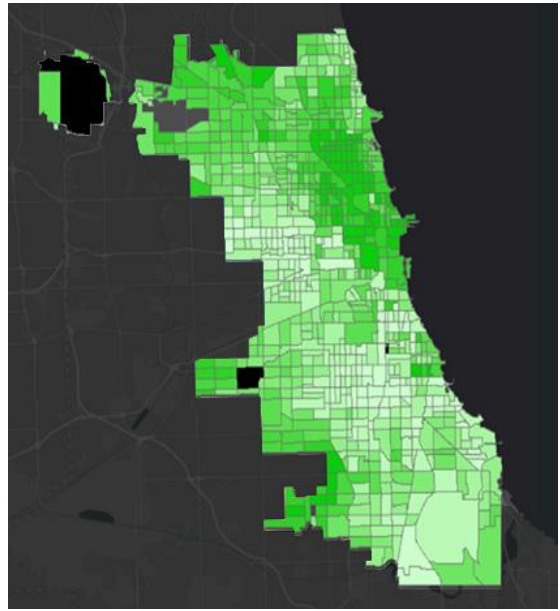

**Figure 2.1:** Parks distribution in Chicago.

**Figure 2.2:** Population density distribution in Chicago.

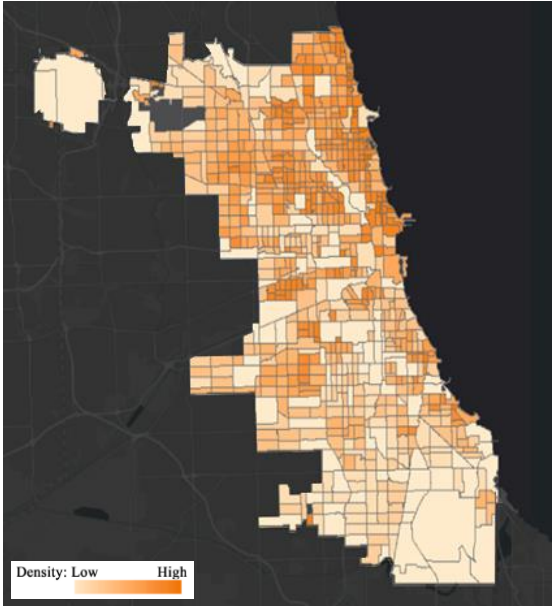

**Figure 2.3:** Spatial distribution of the median income in Chicago.

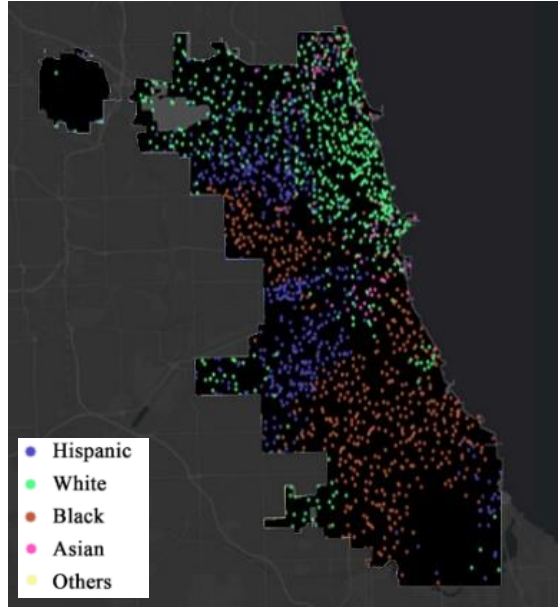

**Figure 2.4:** Spatial distribution of the races in Chicago.

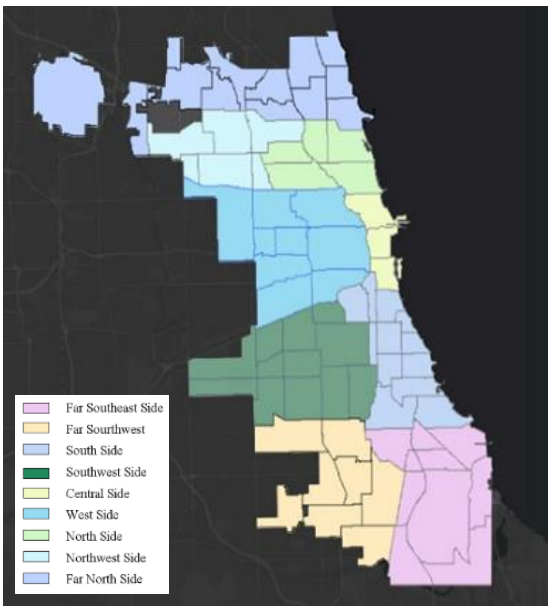

**Figure 2.5:** Community area and sides in Chicago. Chicago is divided into 9 sides and 77 communities

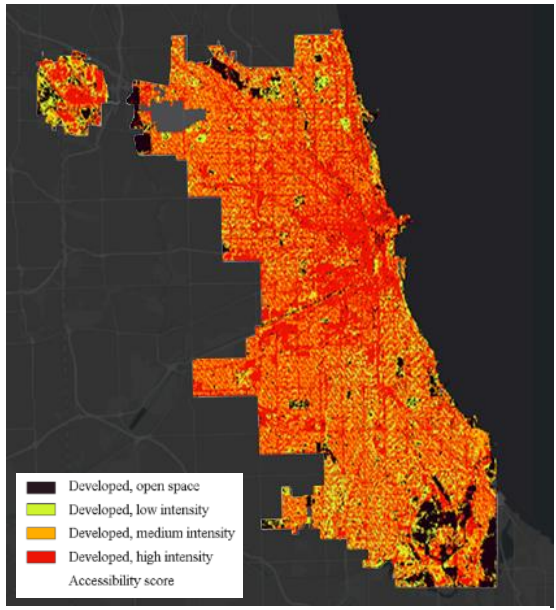

**Figure 2.6:** Landcover map of Chicago.

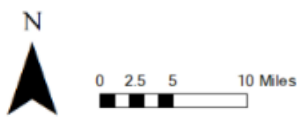

# CHAPTER 3: EVALUATION OF SPATIAL ACCESSIBILITY WORKFLOW

## Definition of the Criteria for the Accessibility Score to Parks

First, all the required data were inputted into ArcGIS 10.7.1. The approach to determine the attractiveness of park is critical to the present research because the present research is about evaluating attractive parks. Individuals go to parks for many different reasons, such as recreation, exercise, socializing with family or friends, watching shows, or breathing fresh air (Neckel et al., 2020). In addition, parks attract visitors for many different reasons, such as a park's high-quality facilities. Parks can be divided into different categories, such as nature parks, recreational parks, sports parks (Egerer et al., 2019). Moreover, each person has their reasons for preferring certain parks over others. A park may be attractive to one group of people but not to others. Consequently, a park's rating and the number of reviews it has received on Google Maps were used to assess its attractiveness. Although people go to parks for different reasons and have various preferences, they were considered to be attracted by a park when they visited it and then rated and reviewed it. Both the rating and review are essential for evaluating the attractiveness of parks. A park was considered attractive when it obtained a high score on Google Maps because this reflected the general attitude of people towards a park Conversely, a low score indicated that people find the park unattractive. Nevertheless, the number of reviews remained essential. The number of reviews a park received reflects the number of people who have visited the park. Consequently, a higher number of reviews attracts more people to a park and increases its attractiveness. The most attractive parks have both a high rating and many reviews. Parks with a low rating and few reviews were considered to be unattractive. However, some parks received a low rating but had many reviews, and these parks were also considered to be attractive because people spent time visiting them and rating them on Google Maps. The following formula was developed to calculate the attractiveness of each park:

$$P_i = R_i * N_i \qquad (1)$$

$P_i$ = the attractiveness score of park i. $R_i$ = the rating of park i, $N_i$ = the number of reviews of park i.

For convenience of the calculation, the 43 parks with no reviews on Google Maps were assigned an attractiveness value of 1, which was the lowest value received by any of the parks.

*Division of research area into land cells*

The landcover data of Chicago was then included in the analysis (Figure 2.6). The landcover data were obtained from the U.S. Geological Survey, and has a resolution of 30 m X 30 m. The land in Chicago is all developed and consists of four types of developed areas: low intensity, medium intensity, high intensity, and open spaces. Regarding the open spaces, the impervious surfaces occupied 0% to 20% of the total landcover. In the low, and medium intensity developed areas the impervious surfaces accounted for 20% to 49%, and 50% to 79% of the total landcover, respectively. Compared to Chicago, in other study areas the impervious surfaces of high intensity developed areas was higher and occupied 80% to 100% of the total landcover (Theobald, 2014). On a map, open space is usually lakes, parks, and woods and is mostly unsuitable for human habitation. The developed area was classified by impervious surface instead of population density. Consequently, it was difficult to determine the relationship between population density and the intensity with which the land had been developed. For example, a residential area with a high-density population, or a huge parking lot are both considered to be high intensity developed areas. In the present study, within one census tract the population was assumed to be evenly distribute in the low, medium, and high intensity developed areas. The areas of open space were removed from the present study because they were unsuitable for people to inhabit.

The analysis of accessibility has gradually become the basis for determining the spatial layout of facilities providing urban public services. To simplify the calculations, the polygons of the parks and the land cells were transformed to their geometric center points in ArcGIS, and these points were used to represent the location of the land cells and parks. Chicago is a large city that consists of a large number of 30 m X 30 m land cells which exceeded the computer's capability. Consequently, the number of 30 m X 30 m land cells were reduced by combining them into 90 m X 90 m land cells.

## The Impedance Factor and Mode of Travel

Walking and driving are the two modes of travel considered in the present research. Because the impedance in the actual transport network should be considered, the measurement of accessibility was based on the road network instead of the straight lines between parks and land cells that would be determined by Euclidean methods. Consequently, measuring distances based on the road network reflects the actual situation and is therefore more accurate. Determining the impedance factor is time consuming. Although walking speed along different types of roads is the same, the driving speed is dissimilar along different types of roads. Consequently, although the distance between two locations is the same, the cost in time to travel between them can be quite different. The application of time costs to measure accessibility is more accurate and precise. Some researchers have indicated that people are willing to travel to parks that take them 30 minutes to reach (Chang and Liao, 2011). Consequently, the maximum time consumed by walking or driving to a park was set at 30 minutes. People walk at an average speed of about 3 mph. Driving speeds are 70 mph on Interstate Highways and US Highways, and 60, 45 and 20 mph on State Highways, Collectors and, Local roads, respectively.

## Measurement of the Cumulative Accessibility Score

The 2SFCA method was applied in calculating the cumulative accessibility score for each land cell (Figure 3.1). The 2SFCA was first proposed in 2000 (Hu et al., 2020) and was given this name in 2003 (Luo & Wang, 2003). The process involved in using it to measure the accessibility of parks is:

1. For a park's central point $i$, all land cell points $j$ within a certain search radius as determined by the time cost $T$, are identified. Then, the score of attraction $S_{ij}$ between the park's central point $i$ and the central point of the land cells $j$ are calculated.

2. For every land cell, all parks $i$ within a certain research radius as determined by the time cost $T$, are identified. The cumulative accessibility $S_j$ is the sum of all the attractions $S_{ji}$ between the parks in the research radius as determined by the time cost $T$, and the land cell $j$.

$$S_i = \sum_{T_{ij} \leq T_0} S_{ij} \quad (2)$$

$$S_j = \sum_{T_{ji} \leq T_0} S_{ji} \quad (3)$$

$S_i$ = the total attractiveness score of a park's central point $i$, $S_j$ = the cumulative score of the attraction of land cell $j$. $S_{ij} = S_{ji}$ = the attraction score between a park's central point $i$ and the central point of a land cell $j$, $T_{ij} = T_{ji}$ = the time costs between a park's central point $i$ and the central point of a land cell $j$ is based on the road network. $T_0$ = maximum time costs.

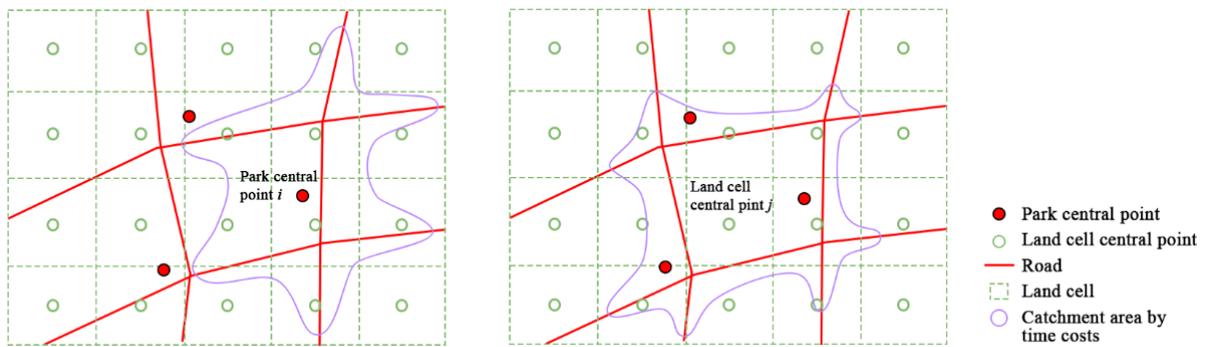

**Figure 3.1:** Process of Two-step floating catchment area (2SFCA) method.

The data of the ratings and reviews of parks has some limitations because certain segments of the population, such as children, elderly or poorly educated people, are not familiar with technical devices and are unable or not willing to rate the parks, which results in errors in the evaluation of the attractiveness of parks (Lee & Hong, 2013). The size of a park is used to evaluate its service capacity. Larger parks are often more attractive than smaller parks because they usually provide larger space and more facilities (Hamilton et al., 1991). The size of the parks was included to reduce these errors and to improve the measurement of the attractiveness of the parks.

## Measurement of the Accessibility Regression

Several regression models have been proposed to measure the regression by time costs (Ma et al., 2018), the gravity model is a theoretical hypothesis that was established by geographers, sociologists, and economists and is based on Newton's universal gravitation formula in classical mechanics, to explain and predict the economic, social, and political interaction of people in geographical space (Baltagi & Egger, 2016). The gravity model is an excellent method to measure the regression of accessibility. In the present research (Figure 3.2a), the attraction score between a park and a land cell was determined by using the park's attractiveness score, the park's size, and the time costs of travelling between the park and the land cell.

$$S_{ji} = S_{ij} = \frac{P_i * A_i}{T_{ij}^2} \qquad (4)$$

$$Sl_j = \sum_{T_{ji} \leq T_0} S_{ji} \qquad (5)$$

$S_{ij}$ = the attraction score between park $i$ and land cell $j$ that can be supplied by land park $i$. $S_{ji}$ = the attraction score between park $i$ and land cell $j$ that can be received by land cell $i$. $Sl_j$ = the cumulative attraction score for land cell $j$ that is within a certain radius based on the time costs, of the parks. $P_i$ = the attractiveness score of park $i$, $A_i$ = the area of park $i$, $T_{ij} = T_{ji}$ the time cost between park $i$ and land cell $j$ based on the road network. $T_0$ = the maximum time costs from land cell $j$ to parks based on the road network.

The linear regression model is another model type used to measure accessibility (Figure 3.2b). The regressive speed is the same from the beginning to the end.

$$S_{ji} = S_{ij} = -\frac{T_{ij}}{T_0} + S_0 \qquad (6)$$

$$Sl_j = \sum_{T_{ji} \leq T_0} S_{ji} \tag{7}$$

$S_0$ = the maximum attraction score between park $i$ and land cell $j$.

The kernel model is similar to the gravity model (Figure 3.2c), and it is also a continuous attenuation function. However, when the time costs are low, the regressive speed of accessibility is slow; whereas the more time is consumed, the faster the regressive speed will be.

$$S_{ji} = S_{ij} = -\frac{3}{4} S_0 \left[1 - \left(\frac{T_{ij}}{T_0}\right)^2\right] \tag{8}$$

$$Sl_j = \sum_{T_{ji} \leq T_0} S_{ji} \tag{9}$$

There are additional regression models, such as the Gaussian model, or piecewise model (Hu et al., 2020). However, only the gravity, linear, and kernel models were used in the present research.

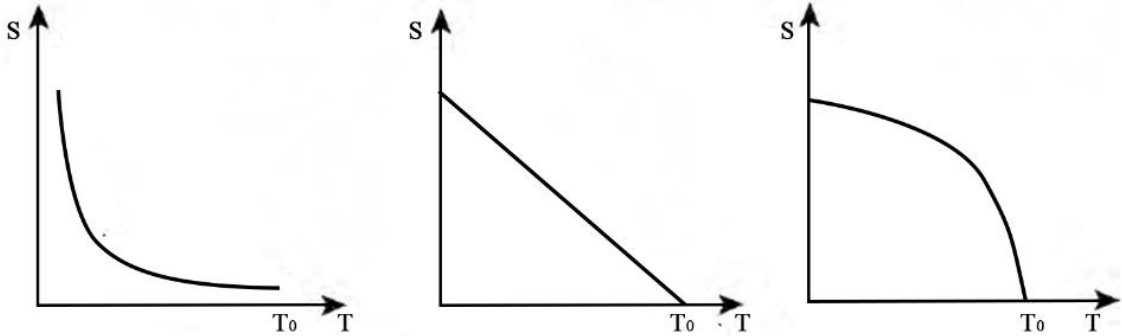

a. Gravity model      b. Linear model      c. Kernel model

**Figure 3.2:** Three regression models. The regression rate for gravity model is fast at the beginning, and slow at the end. The regression rate for kernel model is slow at the beginning, and fast at the end. While the regression rate for linear model keeps the same rate all the time.

Different regression models can be applied to measure the accessibility of parks under different scenarios that are based on people's requirements. When individuals experience an emergency and they need to reach a specific place in the shortest possible time, then the gravity model is better for measuring accessibility. When people are not restricted by time and they want to enjoy the journey, then the kernel model is more suitable. When individuals decide to go to a relatively distant place, walking there will be tiring, but driving will be relaxing. Consequently, the gravity regression model should be considered when walking is involved and the kernel model when driving is involved. In the present study, the gravity, linear, and kernel models were used to calculate the accessibility scores of parks in Chicago by people that were walking and driving. To conveniently and correctly compare the two travel modes using the same regression model, only the gravity model was applied to evaluate the accessibility of parks to different groups of the population.

**Assignment of Census Data to Land Cells**

The land cell data were combined with the census data using spatial join in ArcGIS 10.7.1. The census data include information on the race and income of the people. The population was assumed to be distributed evenly throughout each land cell within a census tract. Moreover, there is no standardized way in which the income per household is classified. Consequently, different studies have used different classifications, and in the present study, household income was classified as follows: low income $0-49999, medium income: $50000- $14999, high income: >$15000. People were assigned to race: White, Black, Hispanic, Asian, and Others. Then, with the accessibility score and census information for each land cell, the spatial distribution of accessibility by different population groups could be calculated using the following formulas:

$$Popl_n = \frac{Popt_m}{N_m} \quad (10)$$

$$Poplg_s = Popl_n * Rlg_s \quad (11)$$

$$Slpc_n = \frac{Sl_n}{Popl_n} \quad (12)$$

$$Slg_s = Sl_n * Rlg_s \quad (13)$$

$Popl_n$ = the population in land cell $n$, $Popt_m$ = the population in census tract $m$, $N_m$ = the number of land cells in census tract $m$. $Poplg_s$ = the population of group $s$ in land cell $n$, $Rlg_s$ = the ratio of the population of groups in land cell $n$. $Slpc_n$ = the score per capita in land cell $n$, $Sl_n$ = the score in land cell $n$. $Slg_s$ = the score for group $s$ in land cell $n$.

## Accessibility at the Community Level

For convenience of comparing the accessibility at the community level, the score of different groups per capita for each community was calculated. The following data were obtained: the people in each land cell, the people in the different groups in each land cell, the accessibility score of parks for the people in each land cell, and the score for different groups in each land cell. The data on the division of the community was combined with the land cell data using spatial join in ArcGIS 10.7.1. The accessibility score at the community level was determined as follows:

$$Scpc_q = \frac{Sc_q}{Popc_q} = \frac{\sum_{i=1}^{k} Sl_n}{\sum_{i=1}^{k} Popl_n} \qquad (14)$$

$$Scgpc_t = \frac{Scg_t}{Popcg_t} = \frac{\sum_{i=1}^{k} Slg_s}{\sum_{i=1}^{k} Poplg_s} \qquad (15)$$

$Scpc_q$ = the accessibility score per capita of community $q$, $Sc_q$ = total accessibility score of the community $q$, $Popc_q$ = the population of community $q$, $k$ = number of land cells in community q, $Scgpc_t$ = the per capita accessibility score for group $t$ in community $q$. $Scg_t$ = the total accessibility score for group $t$ in community $q$.

# CHAPTER 4: THE CUMULATIVE PARK ACCESSIBILITY

## The Cumulative Accessibility by Walking

*The cumulative accessibility by walking for all population groups*

**Figure 4.1:** Total (a-c) and per capita (d-f) cumulative accessibility scores to parks of people walking for 30 minutes. These were calculated using the gravity (a & d), linear (b & e, and kernel (c & f) regression models.

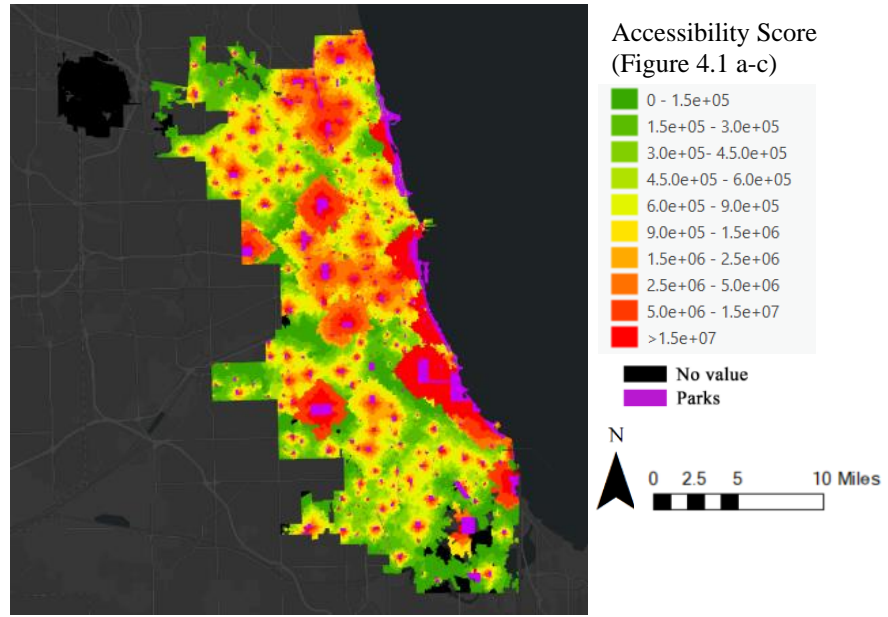

a. Cumulative accessibility score
- gravity model

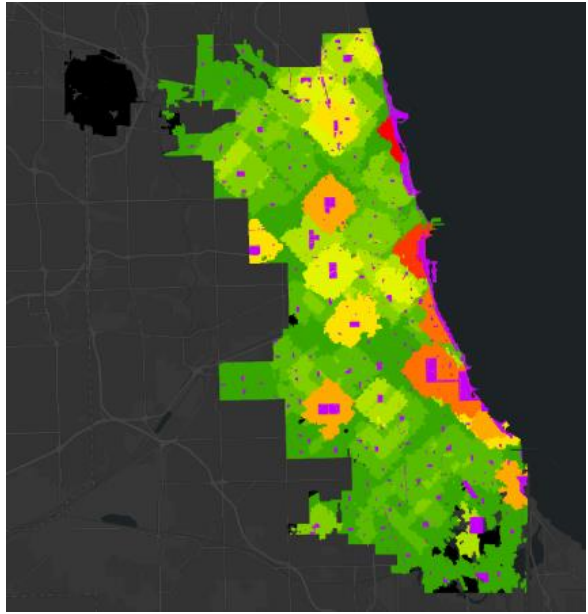

b. Cumulative accessibility score
- linear model

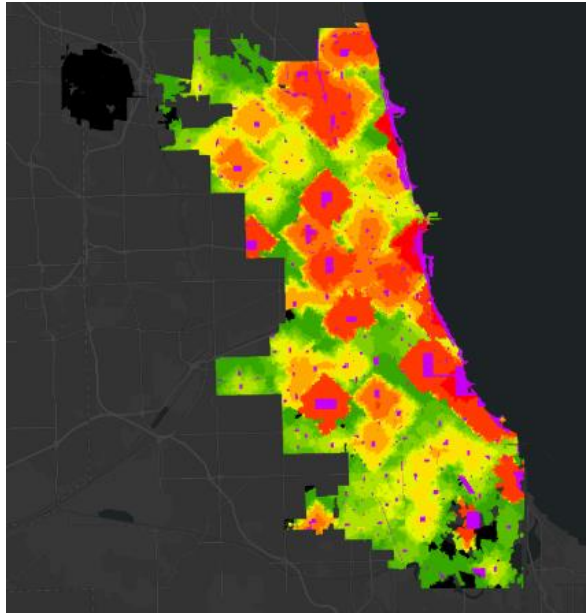

c. Cumulative accessibility score
- kernel model

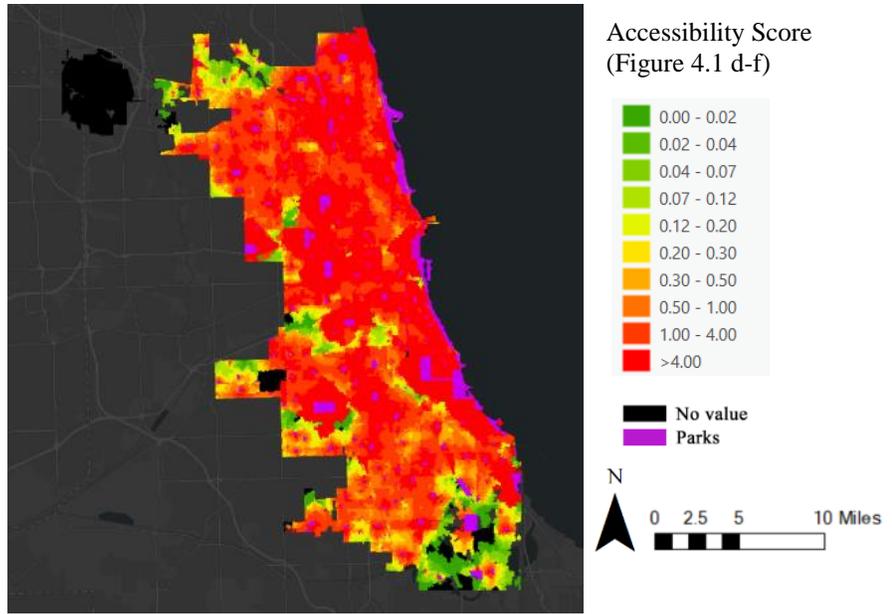

d. Cumulative accessibility score per capita - gravity model

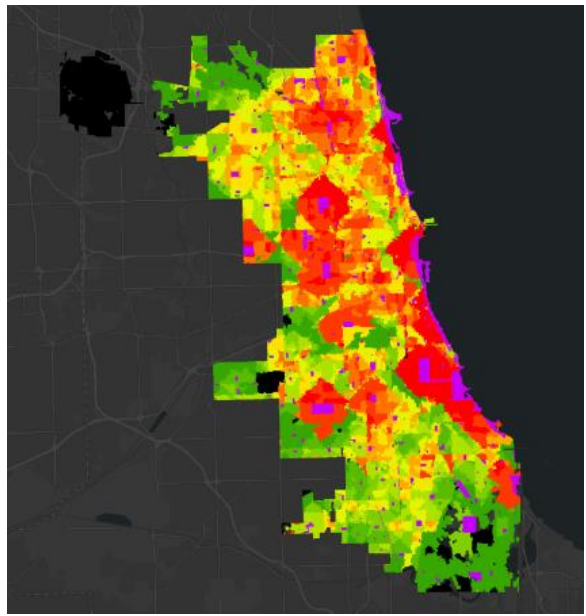

e. Cumulative accessibility score per capita - linear model

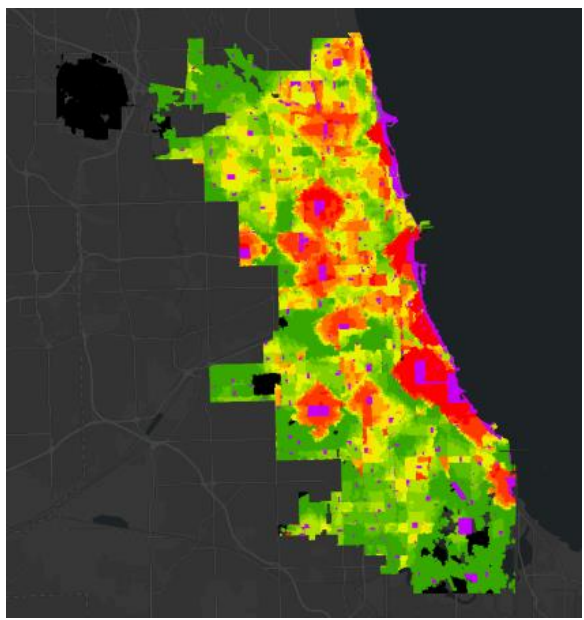

f. Cumulative accessibility score per capita
- kernel model

In Figure 4.1, the black areas within Chicago are open spaces in which people do not live or are not accessible to people walking for 30 minutes. In all three models, certain areas have high accessibility scores to parks, such as the Chicago lakefront, West, and Southwest sides (Figure 2.5, Figure 4.1 a-c). This is because these areas have several attractive parks with high ratings and many reviews, such as Grant Park, Lincoln Park, Washington Park, and Humboldt Park. The standard deviation of the accessibility score to parks determined using the gravity, linear, and kernel model is $1.07 \times 10^{12}$, 7849680, and 3507529, respectively. The standard deviation of the per capita accessibility scores determined using the gravity, linear, and kernel models is 1804958, 107, and 26, respectively. Both lists of standard deviations show that these values from the gravity model are more polarized than those of the other two models. This is because the regression of the gravity model is exponential, which means that when it costs significantly less time to access a park, the accessibility score will be extremely high compared to the accessibility scores of the other two models (Figure 9 a-c). The high accessibility scores to parks on the map of the results of the gravity model are concentrated in the northeast, central and central east, which contrasts with the low scores in the south and northwest of Chicago (Figure 4.1 d). In Figure 4.1 e-f, the per capita accessibility scores determined using the gravity model are much higher than those of the other models. This is because residents have limited access to urban parks by waking for 30 minutes. If people living in a land cell have access to an attractive park relatively quickly, the accessibility score to the park determined using the gravity model should be much higher than those of the other models.

*The cumulative accessibility for different population groups by walking*

**Figure 4.2:** Cumulative accessibility scores to parks of people from different population groups that are walking for 30 minutes

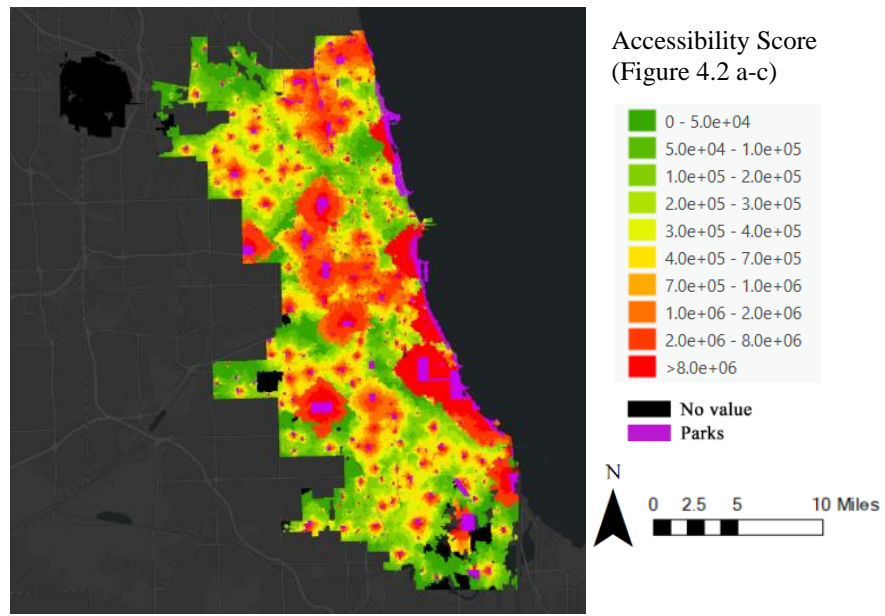

a. Cumulative accessibility score
- low-income population

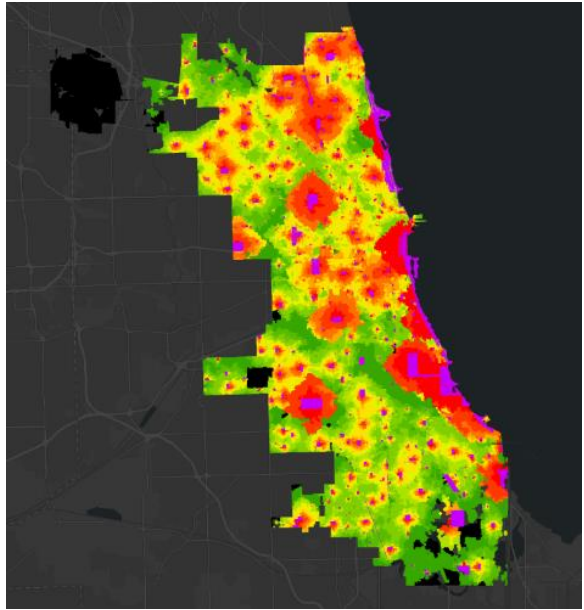

b. Cumulative accessibility score
- median-income population

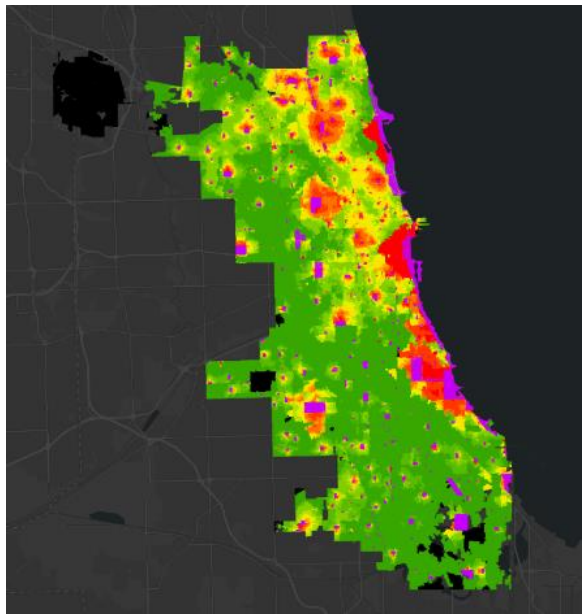

c. Cumulative accessibility score
- high-income population

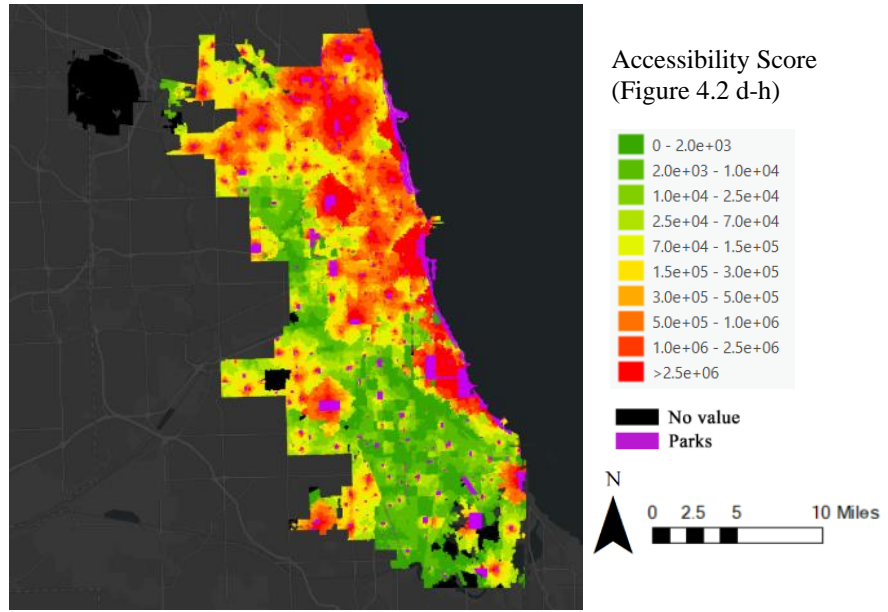

d. Cumulative accessibility score
– the White population

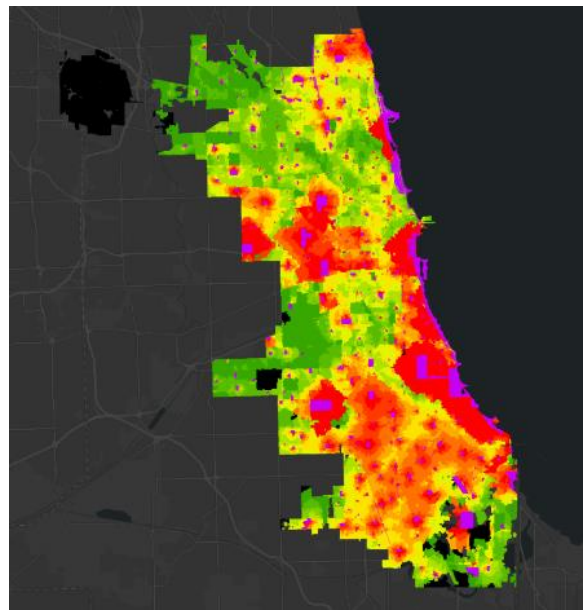

e. Cumulative accessibility score
– the Black population

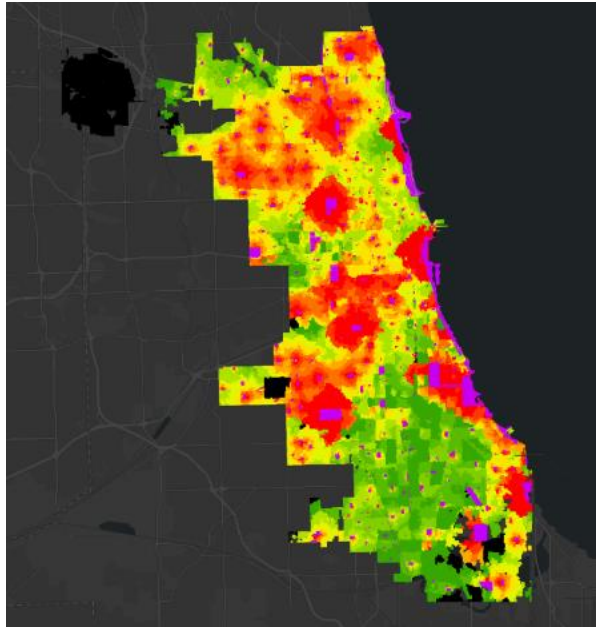

f. Cumulative accessibility score
– the Hispanic population

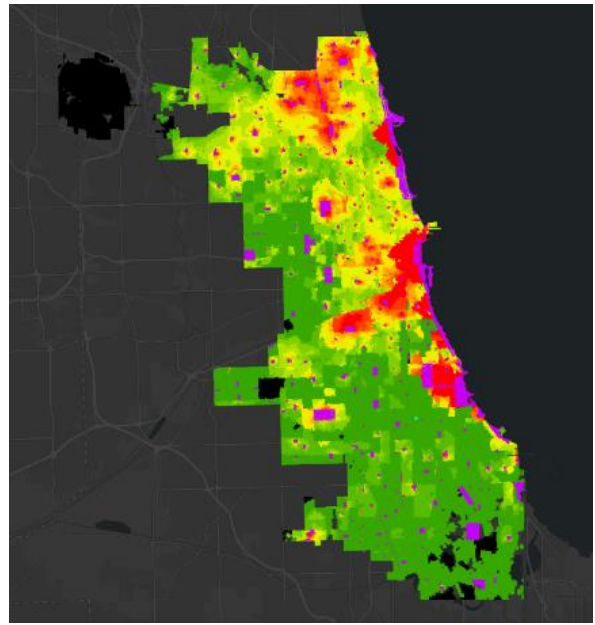

g. Cumulative accessibility score
– the Asian population

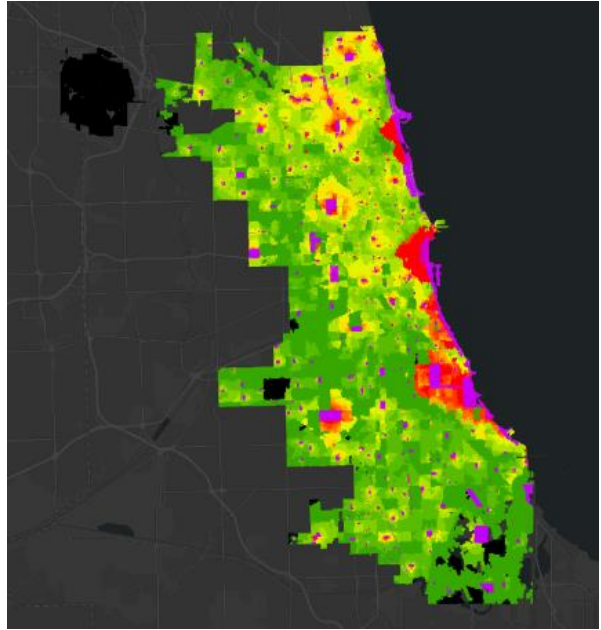

h. Cumulative accessibility score
– other races population

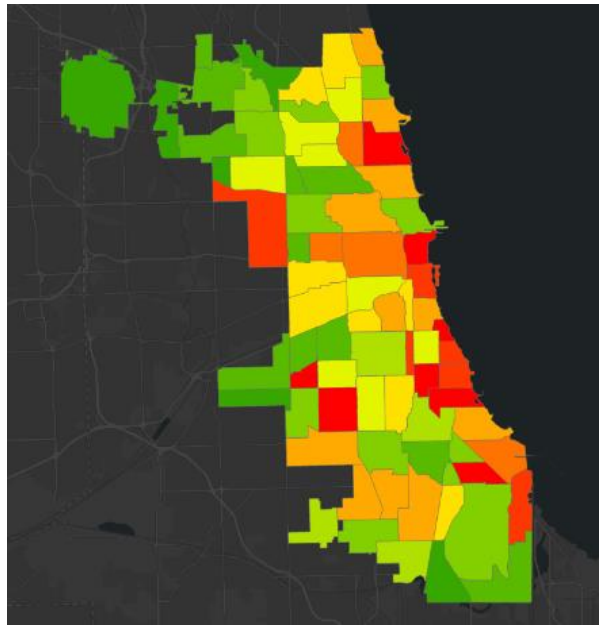
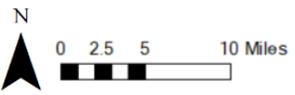

**Figure 4.3:** Cumulative accessibility score - per capita at the community level

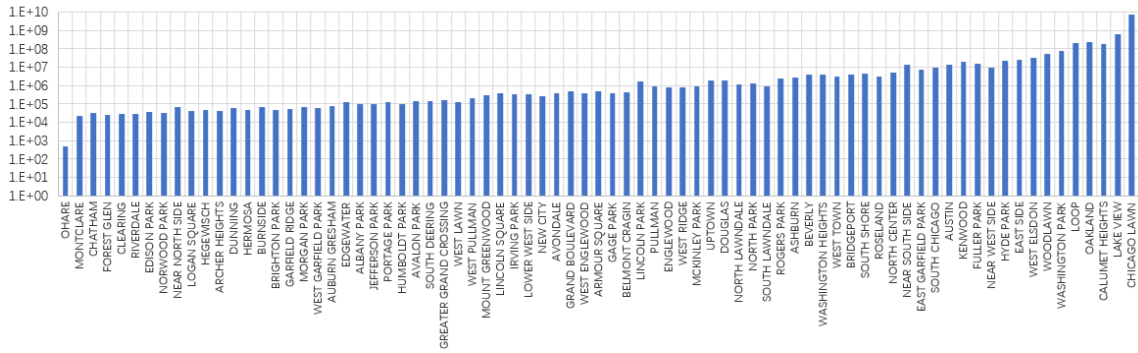

**Figure 4.4:** Value of Community level per capita cumulative accessibility scores to parks by walking

**Table 4.1** Accumulative accessibility score per household for income group at the community level

| Community | Score of low-income population per household | Score of median-income population per household | Score of high-income population per household |
|---|---|---|---|
| OHARE | 1,145 | 1,314 | 1,278 |
| MONTCLARE | 61,267 | 63,723 | 67,145 |
| CHATHAM | 62,473 | 78,683 | 68,652 |
| FOREST GLEN | 66,087 | 64,935 | 74,308 |
| CLEARING | 74,497 | 68,499 | 67,408 |
| RIVERDALE | 77,071 | 63,256 | 74,852 |
| EDISON PARK | 79,498 | 98,692 | 104,755 |
| NORWOOD PARK | 85,132 | 79,956 | 86,003 |
| NEAR NORTH SIDE | 107,210 | 98,165 | 129,714 |
| LOGAN SQUARE | 107,656 | 103,688 | 85,882 |
| HEGEWISCH | 119,506 | 120,822 | 218,652 |
| ARCHER HEIGHTS | 129,014 | 133,763 | 109,525 |
| DUNNING | 140,438 | 172,015 | 169,756 |
| HERMOSA | 145,483 | 159,041 | 120,659 |
| BURNSIDE | 153,106 | 153,106 | / |
| BRIGHTON PARK | 164,526 | 166,670 | 131,783 |
| GARFIELD RIDGE | 166,224 | 147,997 | 141,762 |
| MORGAN PARK | 183,956 | 176,578 | 119,320 |
| WEST GARFIELD PARK | 192,090 | 161,121 | 180,852 |
| AUBURN GRESHAM | 197,598 | 208,459 | 310,208 |
| EDGEWATER | 262,078 | / | 175,929 |
| ALBANY PARK | 278,553 | 295,870 | 297,110 |
| JEFFERSON PARK | 279,185 | 256,729 | 206,424 |

*Table 4.1 (cont.)*

| | | | |
|---|---:|---:|---:|
| PORTAGE PARK | 300,737 | 375,869 | 354,054 |
| HUMBOLDT PARK | 326,648 | 342,701 | 429,686 |
| AVALON PARK | 349,516 | 346,627 | 318,794 |
| SOUTH DEERING | 403,372 | 328,569 | 438,950 |
| GREATER GRAND CROSSING | 408,561 | 412,733 | 165,176 |
| WEST LAWN | 445,312 | 442,351 | 393,507 |
| WEST PULLMAN | 571,447 | 524,181 | 955,004 |
| MOUNT GREENWOOD | 643,407 | 829,160 | 945,908 |
| LINCOLN SQUARE | 780,380 | 843,072 | 844,480 |
| IRVING PARK | 827,185 | 874,713 | 946,693 |
| LOWER WEST SIDE | 857,141 | 868,919 | 865,833 |
| NEW CITY | 917,585 | 748,587 | 781,459 |
| AVONDALE | 989,902 | 1,031,791 | 1,136,583 |
| GRAND BOULEVARD | 1,026,713 | 1,011,154 | 1,112,046 |
| WEST ENGLEWOOD | 1,033,340 | 1,470,296 | 127,405 |
| ARMOUR SQUARE | 1,272,465 | 1,378,221 | 846,084 |
| GAGE PARK | 1,478,310 | 1,437,448 | 2,294,985 |
| BELMONT CRAGIN | 1,539,441 | 1,260,784 | 3,127,863 |
| LINCOLN PARK | 1,745,171 | 3,350,703 | 5,244,501 |
| PULLMAN | 1,829,617 | 2,408,749 | 1,759,993 |
| ENGLEWOOD | 1,942,822 | 2,433,675 | 1,033,224 |
| WEST RIDGE | 2,105,524 | 2,406,119 | 3,432,002 |
| MCKINLEY PARK | 2,431,514 | 3,330,416 | 2,414,307 |
| UPTOWN | 2,700,139 | 4,566,051 | 5,443,495 |
| DOUGLAS | 3,742,033 | 4,679,292 | 2,317,291 |
| NORTH LAWNDALE | 4,080,617 | 2,709,603 | 7,397,216 |
| NORTH PARK | 4,153,718 | 3,146,324 | 4,205,410 |
| SOUTH LAWNDALE | 4,247,119 | 3,386,551 | 2,754,746 |
| ROGERS PARK | 5,318,850 | 5,287,526 | 4,618,887 |
| ASHBURN | 6,301,688 | 9,801,332 | 14,847,932 |
| BEVERLY | 7,044,580 | 12,146,660 | 9,529,841 |
| WASHINGTON HEIGHTS | 8,278,309 | 14,131,984 | 13,997,785 |
| WEST TOWN | 8,820,419 | 5,553,841 | 6,803,478 |
| BRIDGEPORT | 9,064,741 | 12,176,653 | 8,778,321 |
| SOUTH SHORE | 10,911,081 | 8,491,787 | 6,820,064 |
| ROSELAND | 11,857,675 | 6,150,598 | 241,413 |
| NORTH CENTER | 16,123,052 | 10,724,073 | 13,031,883 |
| NEAR SOUTH SIDE | 23,200,244 | 25,557,177 | 24,883,650 |
| EAST GARFIELD PARK | 24,780,964 | 10,051,669 | 452,319 |
| SOUTH CHICAGO | 25,104,969 | 27,304,807 | 20,398,772 |
| AUSTIN | 25,615,455 | 52,915,962 | 69,928,103 |
| KENWOOD | 32,599,959 | 38,038,560 | 58,082,654 |
| FULLER PARK | 33,708,323 | 31,406,483 | / |

*Table 4.1 (cont.)*

| | | | |
|---|---|---|---|
| NEAR WEST SIDE | 46,305,545 | 10,600,809 | 5,354,825 |
| HYDE PARK | 49,248,886 | 44,503,410 | 32,641,000 |
| EAST SIDE | 85,679,193 | 85,230,066 | 35,414,787 |
| WEST ELSDON | 128,300,623 | 124,753,549 | 133,861,554 |
| WOODLAWN | 142,444,097 | 99,033,824 | 84,795,930 |
| WASHINGTON PARK | 166,267,083 | 205,532,272 | 228,378,145 |
| LOOP | 255,864,451 | 246,598,578 | 592,883,077 |
| OAKLAND | 480,610,092 | 575,533,503 | 669,684,551 |
| CALUMET HEIGHTS | 516,540,845 | 433,396,219 | 315,264,005 |
| LAKE VIEW | 1,475,924,522 | 1,116,781,284 | 868,626,093 |
| CHICAGO LAWN | 26,010,156,903 | 17,487,963,292 | 23,545,224,450 |

**Table 4.2:** Accumulative accessibility score per capita for race groups by walking at the community level

| Community | Score of white population per capita | Score of black population per capita | Score of Hispanic population per capita | Score of Asian population per capita | Score of other races population per capita |
|---|---|---|---|---|---|
| OHARE | 548 | 842 | 165 | 1,084 | 958 |
| MONTCLARE | 19,909 | 28,133 | 21,850 | 27,755 | 21,181 |
| CHATHAM | 15,137 | 30,727 | 18,328 | 44,670 | 20,693 |
| FOREST GLEN | 25,107 | 17,107 | 23,947 | 28,372 | 27,132 |
| CLEARING | 21,820 | 27,318 | 26,516 | 9,327 | 16,263 |
| RIVERDALE | 39,957 | 26,764 | 16,876 | 13,880 | 21,442 |
| EDISON PARK | 37,723 | 51,347 | 43,898 | 12,533 | 27,440 |
| NORWOOD PARK | 33,774 | 58,799 | 29,992 | 21,081 | 46,189 |
| NEAR NORTH SIDE | 64,831 | 42,128 | 64,910 | 105,896 | 80,759 |
| LOGAN SQUARE | 38,126 | 46,944 | 44,255 | 33,236 | 38,067 |
| HEGEWISCH | 55,662 | 7,509 | 58,879 | 543,184 | 12,276 |
| ARCHER HEIGHTS | 38,814 | 49,302 | 38,977 | 34,416 | 31,122 |
| DUNNING | 53,988 | 62,315 | 62,236 | 66,790 | 49,454 |
| HERMOSA | 53,480 | 46,267 | 43,273 | 35,338 | 70,291 |
| BURNSIDE | 68,557 | 68,557 | 68,557 | / | 68,557 |
| BRIGHTON PARK | 42,108 | 86,450 | 41,823 | 76,702 | 83,564 |
| GARFIELD RIDGE | 43,168 | 179,445 | 47,161 | 54,341 | 83,501 |
| MORGAN PARK | 47,381 | 74,885 | 27,528 | 8,121 | 120,813 |
| WEST GARFIELD PARK | 129,217 | 57,199 | 40,979 | 14,351 | 57,216 |
| AUBURN GRESHAM | 56,596 | 76,642 | 93,186 | 81,451 | 46,651 |
| EDGEWATER | 121,810 | 124,450 | 85,097 | 123,942 | 275,744 |
| ALBANY PARK | 118,262 | 126,714 | 79,131 | 87,886 | 117,017 |
| JEFFERSON PARK | 92,586 | 189,326 | 100,767 | 82,228 | 144,465 |
| PORTAGE PARK | 135,237 | 179,322 | 100,244 | 96,032 | 167,513 |

*Table 4.2 (cont.)*

| | | | | | |
|---|---|---|---|---|---|
| HUMBOLDT PARK | 178,805 | 108,712 | 89,947 | 111,349 | 70,692 |
| AVALON PARK | 159,802 | 138,391 | 198,688 | 204,501 | 99,029 |
| SOUTH DEERING | 154,694 | 107,704 | 185,247 | 38,482 | 99,603 |
| GREATER GRAND CROSSING | 311,397 | 159,113 | 107,493 | 81,056 | 206,646 |
| WEST LAWN | 96,018 | 174,666 | 129,523 | 190,372 | 121,711 |
| WEST PULLMAN | 40,009 | 204,180 | 50,934 | 10,473 | 269,463 |
| MOUNT GREENWOOD | 300,199 | 318,603 | 182,388 | 645,572 | 348,013 |
| LINCOLN SQUARE | 367,995 | 265,488 | 392,080 | 293,867 | 341,808 |
| IRVING PARK | 325,957 | 263,630 | 320,631 | 372,340 | 424,095 |
| LOWER WEST SIDE | 278,619 | 192,697 | 338,554 | 303,825 | 366,465 |
| NEW CITY | 111,009 | 731,769 | 129,560 | 121,877 | 571,407 |
| AVONDALE | 434,372 | 437,945 | 301,600 | 651,496 | 275,633 |
| GRAND BOULEVARD | 521,826 | 467,751 | 538,934 | 481,667 | 439,857 |
| WEST ENGLEWOOD | 219,683 | 326,896 | 915,986 | 31,500 | 460,239 |
| ARMOUR SQUARE | 276,936 | 216,903 | 329,252 | 612,145 | 237,926 |
| GAGE PARK | 211,877 | 601,154 | 369,243 | 181,066 | 767,048 |
| BELMONT CRAGIN | 212,653 | 800,650 | 451,270 | 130,280 | 33,193 |
| LINCOLN PARK | 1,827,442 | 195,566 | 2,575,393 | 1,417,898 | 409,715 |
| PULLMAN | 109,821 | 1,049,771 | 333,649 | 50,491 | 78,257 |
| ENGLEWOOD | 988,315 | 774,476 | 779,985 | 49,182 | 922,813 |
| WEST RIDGE | 1,002,558 | 485,065 | 772,120 | 666,012 | 245,279 |
| MCKINLEY PARK | 919,130 | 1,255,888 | 802,882 | 1,192,325 | 181,341 |
| UPTOWN | 2,458,705 | 751,127 | 1,370,160 | 1,939,407 | 1,767,739 |
| DOUGLAS | 707,380 | 2,007,044 | 602,188 | 1,803,768 | 2,148,667 |
| NORTH LAWNDALE | 1,327,579 | 1,560,235 | 161,204 | 116,428 | 3,537,170 |
| NORTH PARK | 1,486,719 | 1,309,667 | 723,017 | 1,474,183 | 846,176 |
| SOUTH LAWNDALE | 1,963,164 | 4,781,703 | 323,209 | 1,418,448 | 47,041 |
| ROGERS PARK | 3,187,418 | 2,031,179 | 1,058,435 | 2,036,913 | 1,571,501 |
| ASHBURN | 883,775 | 5,944,784 | 171,064 | 27,127 | 2,306,473 |
| BEVERLY | 5,531,498 | 1,001,924 | 2,683,348 | 651,111 | 3,014,553 |
| WASHINGTON HEIGHTS | 1,725,575 | 3,905,781 | 2,797,946 | 182,905 | 6,504,830 |
| WEST TOWN | 2,639,044 | 3,998,730 | 3,218,893 | 4,197,912 | 5,036,738 |
| BRIDGEPORT | 2,585,002 | 2,941,995 | 2,750,395 | 5,709,951 | 5,940,862 |
| SOUTH SHORE | 3,162,677 | 4,361,528 | 6,057,723 | 18,700,957 | 7,944,271 |
| ROSELAND | 5,591,135 | 3,199,731 | 5,325,422 | 126,306 | 2,054,568 |
| NORTH CENTER | 3,813,007 | 6,802,531 | 11,248,109 | 7,422,508 | 6,108,799 |
| NEAR SOUTH SIDE | 14,396,900 | 13,273,734 | 15,220,048 | 14,185,973 | 13,788,546 |
| EAST GARFIELD PARK | 7,946,668 | 7,397,861 | 3,992,432 | 344,403 | 117,896 |
| SOUTH CHICAGO | 4,308,216 | 10,045,693 | 5,998,586 | 45,293,606 | 17,558,215 |
| AUSTIN | 37,793,134 | 7,728,801 | 14,969,099 | 37,372,392 | 31,315,859 |
| KENWOOD | 51,337,634 | 12,624,019 | 18,346,637 | 9,086,902 | 31,580,092 |
| FULLER PARK | 15,485,305 | 15,754,140 | 8,914,147 | 641,227 | 641,227 |
| NEAR WEST SIDE | 2,790,289 | 25,653,923 | 5,897,446 | 2,273,180 | 14,191,810 |

*Table 4.2 (cont.)*

| | | | | | |
|---|---|---|---|---|---|
| HYDE PARK | 14,773,598 | 31,267,115 | 29,009,736 | 17,285,467 | 21,902,461 |
| EAST SIDE | 11,460,907 | 25,521,619 | 26,087,771 | 178,002 | 18,105 |
| WEST ELSDON | 44,101,170 | 71,418,643 | 32,242,986 | 36,226 | 57,335,917 |
| WOODLAWN | 25,839,351 | 55,929,513 | 34,694,526 | 35,871,798 | 145,033,743 |
| WASHINGTON PARK | 111,117,218 | 69,230,209 | 208,335,576 | 259,014,154 | 22,414,142 |
| LOOP | 199,445,723 | 220,516,179 | 187,401,619 | 176,390,436 | 268,959,284 |
| OAKLAND | 256,953,994 | 229,237,758 | 314,105,219 | 334,032,417 | 224,301,216 |
| CALUMET HEIGHTS | 223,437,123 | 175,265,726 | 604,303,426 | 460,769,638 | 54,519 |
| LAKE VIEW | 625,344,968 | 390,715,785 | 225,294,433 | 895,652,241 | 532,137,766 |
| CHICAGO LAWN | 4,190,391,728 | 13,079,330,682 | 1,780,313,782 | 12,678,280,865 | 28,866,224,089 |

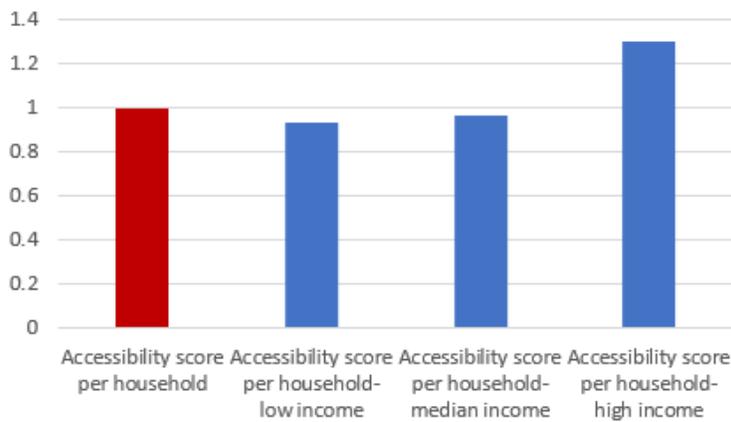

a. Cumulative accessibility scores per household
cumulative accessibility scores

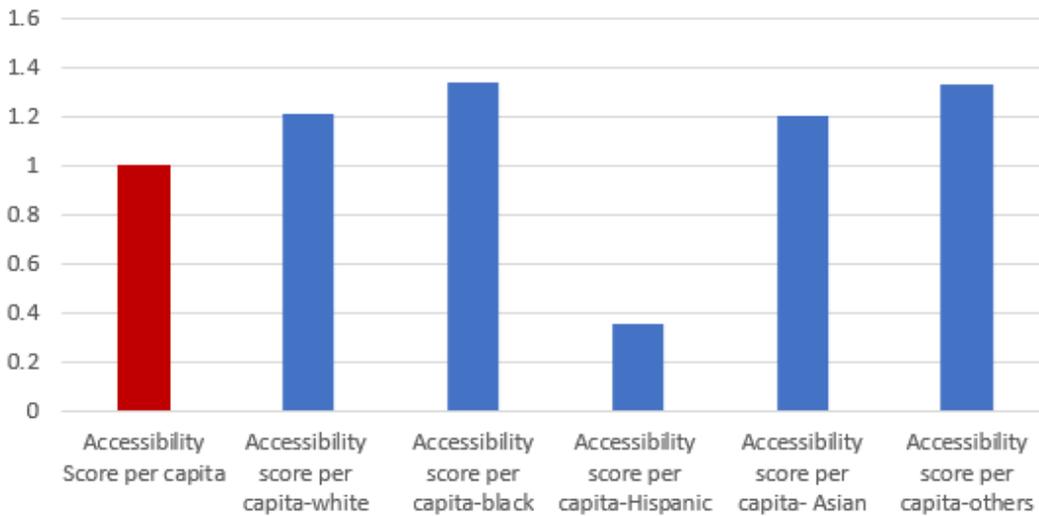

b. Per capita cumulative accessibility scores for different race groups

**Figure 4.5:** Per capita cumulative accessibility scores to parks by walking of people from different population groups.

In this paper, the gravity model was used to determine and evaluate the accessibility scores of different groupings of people. Figure 4.2 a-c shows the spatial distribution of accessibility to parks of people with three different levels of income. This accessibility score was determined using the population of a certain group within each land cell, and the score of accessibility within each land cell. On all three maps (Figure 4.2 a-c), for people of all three levels of income the scores are high around the attractive parks. The distribution of the accessibility scores depicted on the three maps corresponds with the distribution of the median income depicted on the map of Chicago in Figure 2.3. The accessibility score of the high-income population in the Central and North Sides of Chicago is very high because the median income in these areas is high, which means that relatively wealthy segments of the population are concentrated in these areas. The accessibility score of the low-income population is high in the West, Southwest, and Far Southwest Sides. These results also correspond with the distribution of median income depicted on the map in Figure 2.3. The median income in these areas is low, which means that many low-income populations live there. Figure 4.2 d-h, illustrates the accessibility scores to parks of people from different racial groups. Similar to the accessibility scores illustrated on the maps of people with three different levels of income, they correspond with the attractive parks and the distribution of the different population groups. The White population has high accessibility scores to parks in the Far North, North, and Central Sides of Chicago. The Black population has high accessibility scores in the Southwest, South, and Far Southwest Sides of Chicago. The Hispanic population has high accessibility scores in the West, Southwest, and Northwest Sides. The Asian population has high accessibility scores to parks in the Central, South, and Far North

Sides. The other races have high accessibility scores to parks in the South, and Central Sides of Chicago.

*The cumulative accessibility at community level by walking*

Figure 4.3 and Figure 4.4 show the community level per capita accessibility score of people that are walking. Even though they are from the same community, people from different population groups have different per capita accessibility scores to parks. This is because the per capita accessibility scores of the different population groups in every land cell is the same. However, the community level per capita accessibility scores to parks are calculated by dividing the total accessibility score of the different groups in all land cells within a community by the total number of populations of the different groups. There are two variants, so the results are different. Several communities along the Chicago lakefront and in the west have high accessibility scores to parks, such as the Chicago Lawn and Lake View community. For convenience of visualization, the logarithmic scale is used to show the values of the scores. Table 4.1 and Table 4.2 shows the community level per household or per capita accessibility scores of different population groups to parks of people walking. The accessibility scores to parks for each group of people within every land cell is the same. However, when the land cells are combined and considered as a community the score of accessibility for every individual in one community will be different. Because of the application of the gravity regression model, the difference between different communities is enormous. Considering the accessibility scores to parks of different races and levels of income, the O'Hare community has the worst accessibility to parks, and the Chicago Lawn Community has the best accessibility to parks. Within a community the difference in the accessibility to parks was greater between the different racial groups than between people with different levels of income.

Figure 4.5 shows the accessibility score per household and per capita of different population groups. The high-income population has the highest accessibility to parks compared to the other two groups. The accessibility scores to parks of the low- and median income populations is below the average level (Figure 4.5a). Comparing between the races, the Black population has the best access to attractive parks, whereas the Hispanic population has the least access (Figure 4.5b). This is because most of the Hispanic population is distributed far from the many highly attractive parks in the lakefront area (Figure 2.4). Most Hispanic people do not have access to attractive parks by walking for 30 minutes.

## The Cumulative Accessibility by Driving

*The cumulative accessibility by driving for all population groups*

**Figure 4.6:** Cumulative accessibility scores to parks for people driving for 30 minutes.

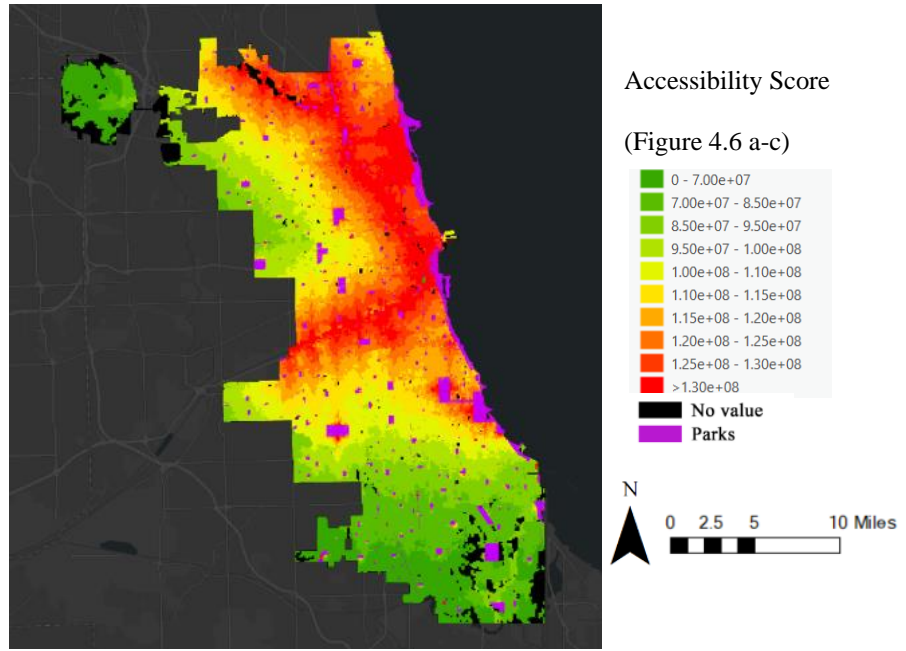

a. Cumulative accessibility score
- gravity model

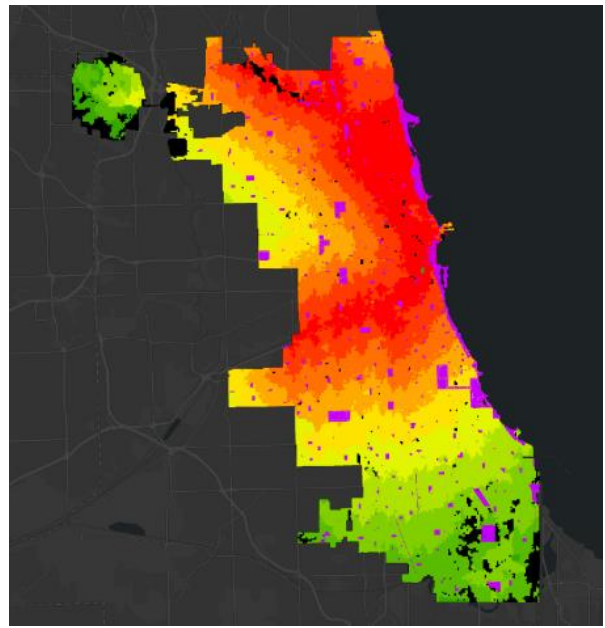

b. Cumulative accessibility score
- linear model

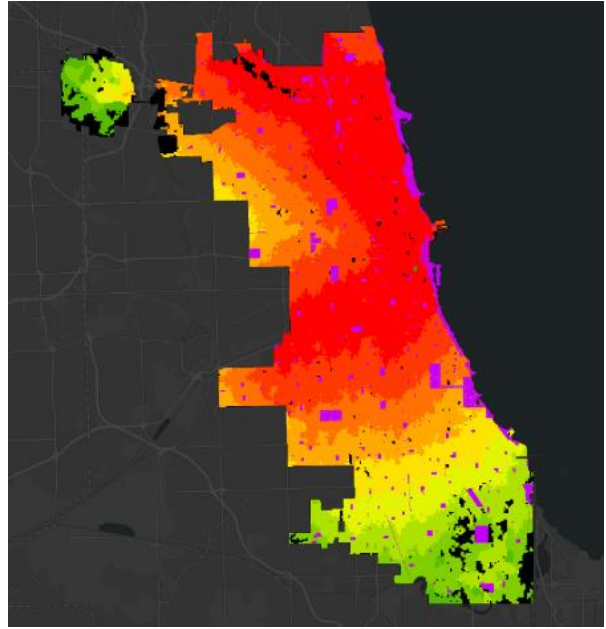

c. Cumulative accessibility score
- kernel model

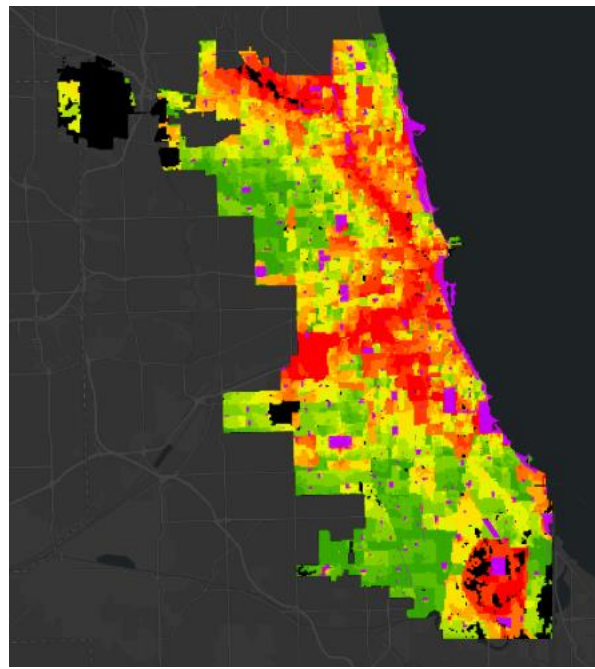
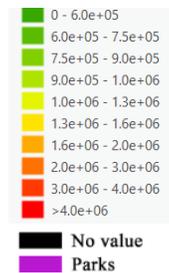
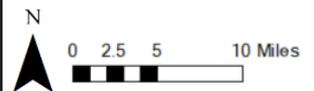

d. Cumulative accessibility score
- per capita gravity model

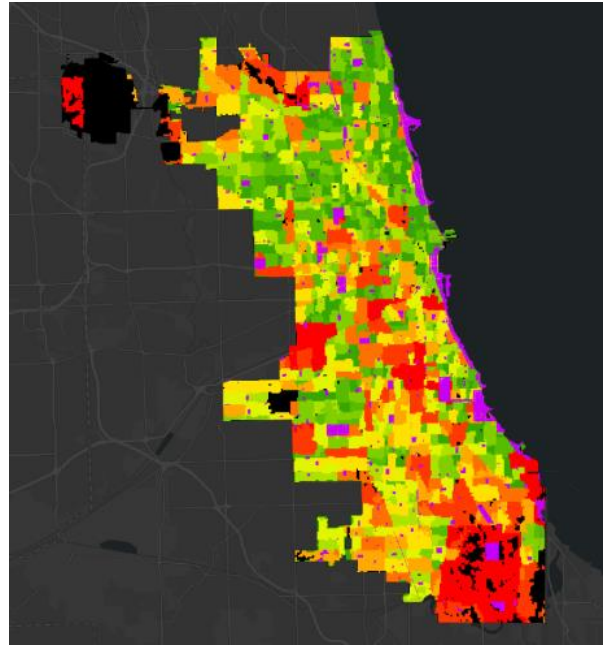

e. Cumulative accessibility score
- per capita linear model

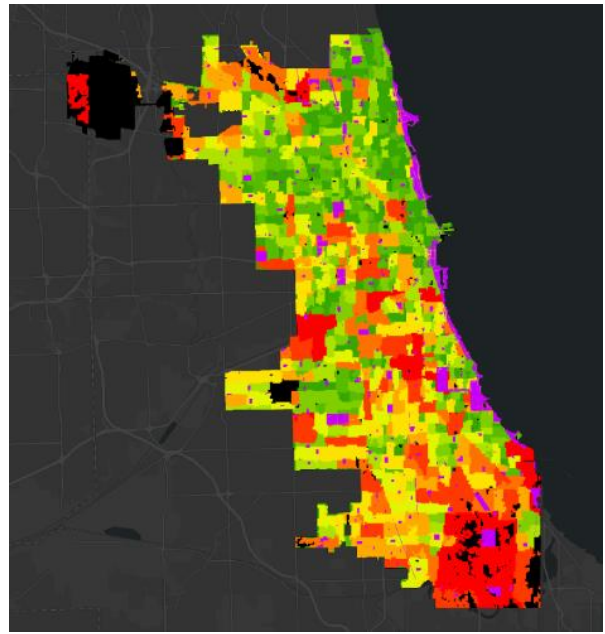

f. Cumulative accessibility score
- per capita kernel model

Figure 4.6 a-c, shows the cumulative accessibility score to parks of people diving for 30 minutes. The scores were determined using the gravity, linear, and kernel regression models. The areas with high accessibility scores are along Interstates 90 and 55 which bisect Chicago on all three maps. The results of the three models are similar. However, because the regression of the gravity model is much faster than the other two models, the area with a high accessibility score is a narrow band either side of the major roads (Figure 4.6a). In contrast to this, because the regression for the kernel model is slower than the other two models, the area with a high accessibility score is a broad band either side of the major roads (Figure 4.6c). Compared with the other two models, the linear model has a linear regression and is more stable. Consequently, the area with a high accessibility score estimated by the linear model, is a band either side of the major roads which is intermediate between the bands predicted by the other two models (Figure 4.6b). Figure 4.6 a-c also reflect the tendency of the accessibility scores to parks of the gravity model being more concentrated. Figure 4.6 d-f, show the per capita cumulative accessibility scores to parks by people driving for 30 minutes as estimated using the gravity, linear, and kernel regression models. All three models estimate a higher per capita accessibility score to parks of people in the Far Southeast Side of Chicago. Even though the accessibility score is not high, because the population is small, when dividing the accessibility score by the small population it results in the per capita accessibility score to parks being higher.

*The cumulative accessibility by driving for different population groups*

**Figure 4.7:** Cumulative accessibility score by driving for 30 minutes of different population groups

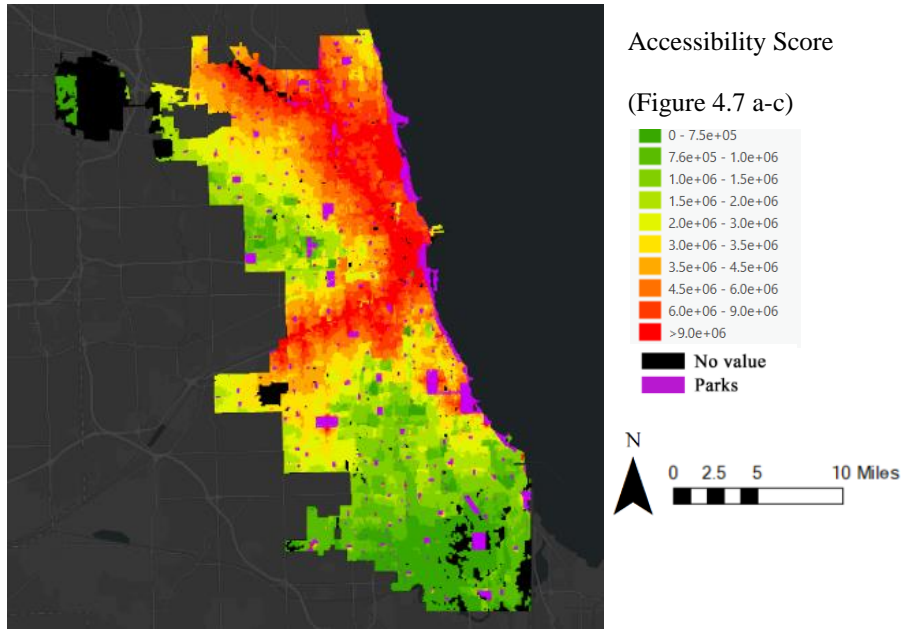

a. Cumulative accessibility score
- low-income population

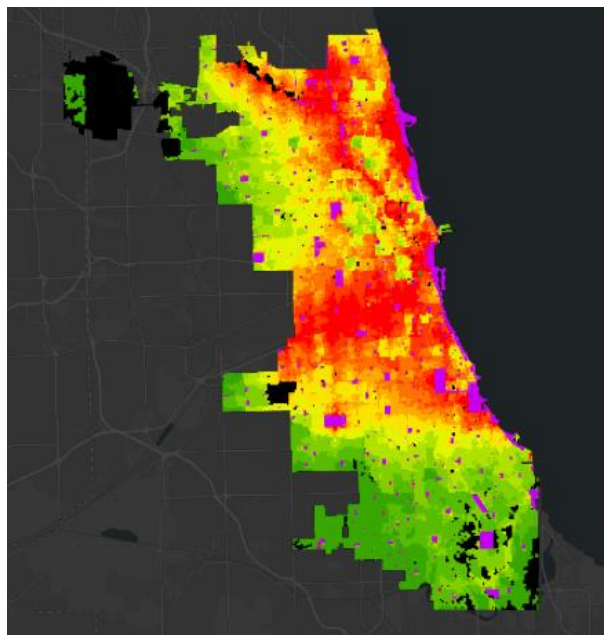

b. Cumulative accessibility score
- median-income population

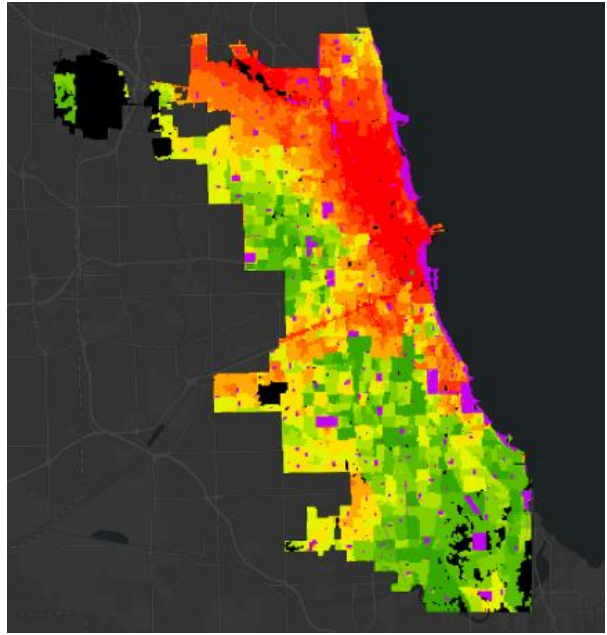

c. Cumulative accessibility score
 - high-income population

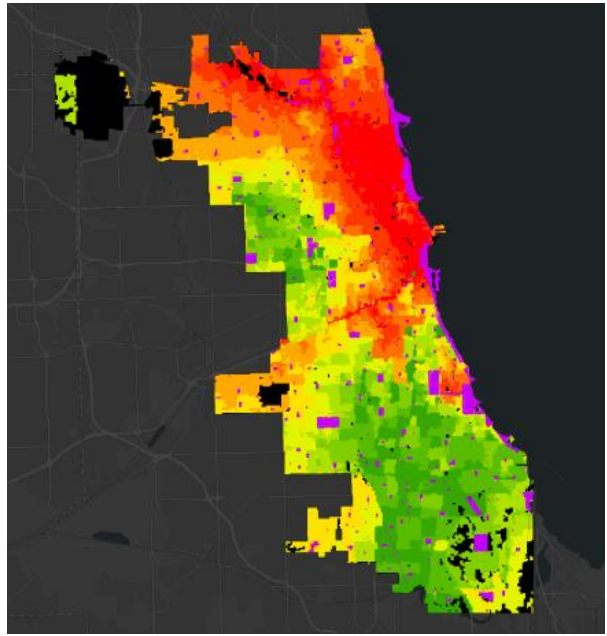

d. Cumulative accessibility score
 – the White population

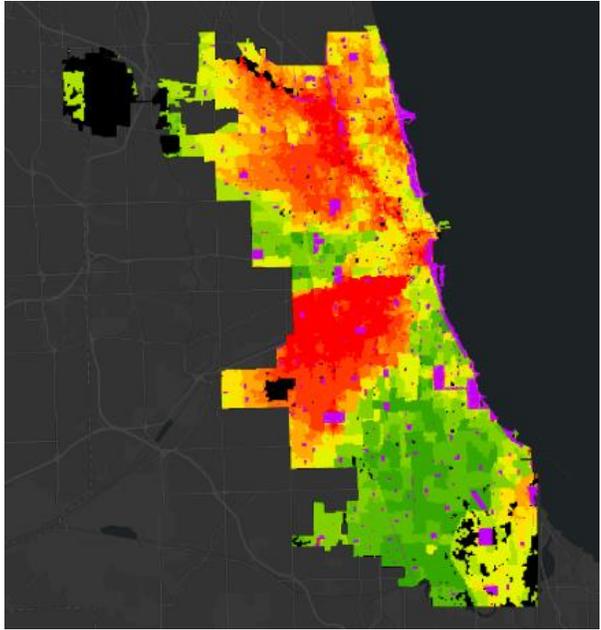

e. Cumulative accessibility score
– the Black population

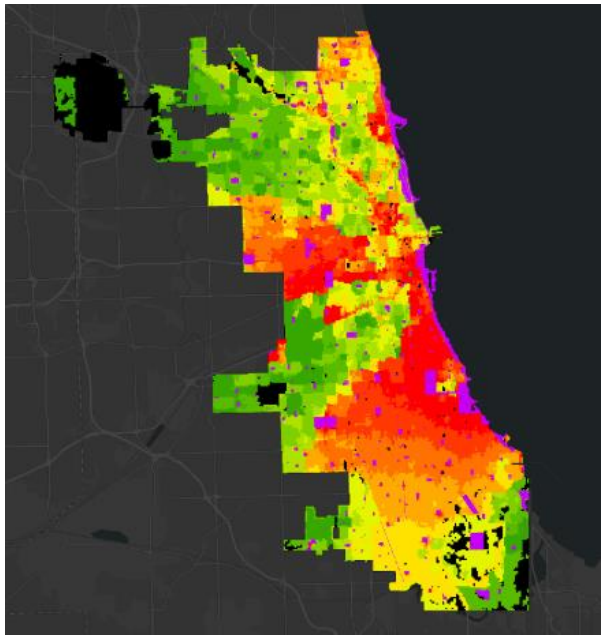

f. Cumulative accessibility score – the Hispanic population

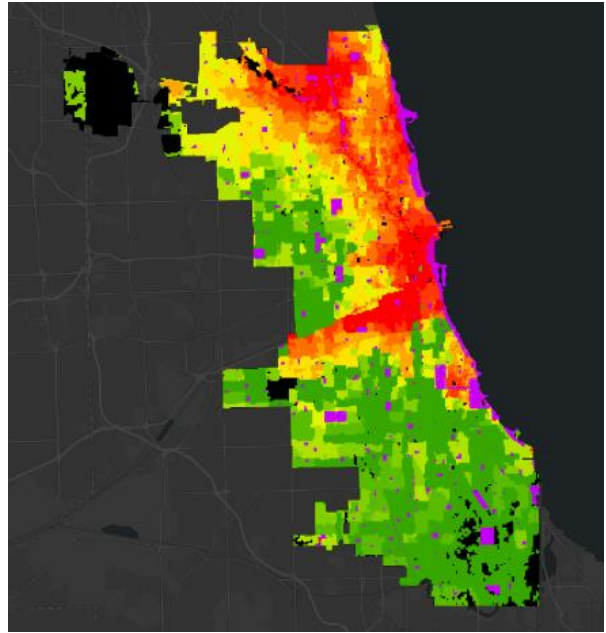

g. Cumulative accessibility score
– the Asian population

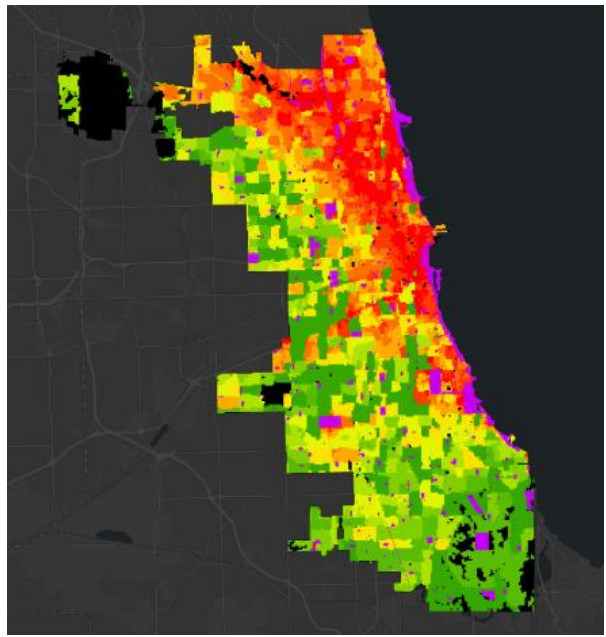

h. Cumulative accessibility score
– other races population

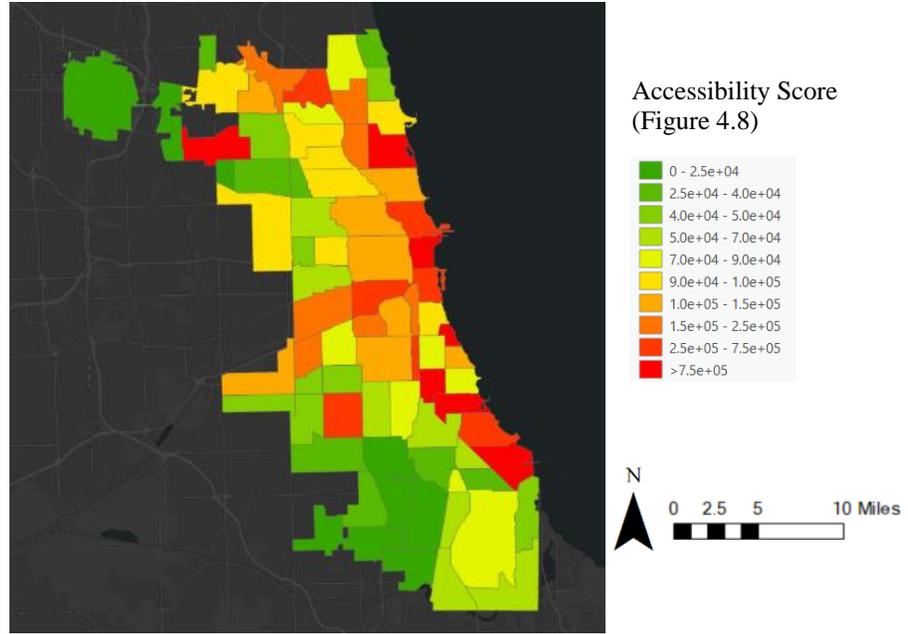

**Figure 4.8:** Cumulative accessibility score
- per capita at the community level

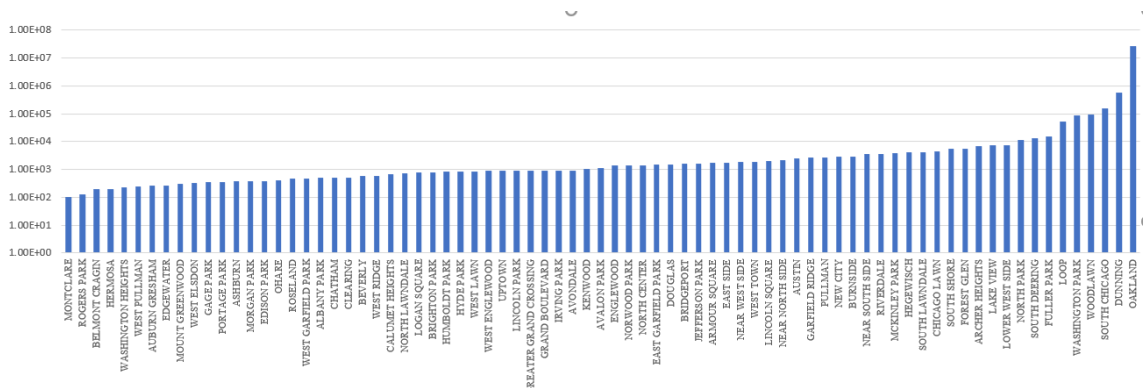

**Figure 4.9:** Value of Community level per capita cumulative accessibility scores to parks by driving

**Table 4.3:** Accumulative accessibility score per household for income groups by driving at the community level

| Community | Score of low-income population per household | Score of median-income population per household | Score of high-income population per household |
|---|---|---|---|
| MOUNT GREENWOOD | 9,580 | 24,502 | 12,707 |
| OHARE | 3,237 | 17,667 | 15,580 |
| MONTCLARE | 3,413 | 15,683 | 19,286 |

*Table 4.3 (cont.)*

| | | | |
|---|---|---|---|
| BEVERLY | 25,925 | 35,348 | 19,483 |
| WEST PULLMAN | 2,004 | 14,826 | 22,770 |
| MORGAN PARK | 7,568 | 24,184 | 25,109 |
| EDISON PARK | 34,839 | 46,962 | 25,951 |
| WASHINGTON HEIGHTS | 3,899 | 25,735 | 28,796 |
| ASHBURN | 9,511 | 48,999 | 29,855 |
| ROGERS PARK | 4,610 | 22,235 | 33,493 |
| CHATHAM | 2,097 | 25,373 | 40,543 |
| AUBURN GRESHAM | 2,693 | 21,425 | 41,357 |
| EDGEWATER | 13,543 | 39,510 | 43,270 |
| ROSELAND | 2,861 | 26,890 | 45,686 |
| PORTAGE PARK | 18,507 | 63,080 | 46,735 |
| CALUMET HEIGHTS | 6,908 | 39,183 | 47,810 |
| CLEARING | 9,413 | 59,968 | 49,788 |
| BELMONT CRAGIN | 8,642 | 47,322 | 55,543 |
| HERMOSA | 6,816 | 38,205 | 56,743 |
| LOGAN SQUARE | 66,904 | 100,267 | 59,493 |
| LINCOLN PARK | 119,578 | 126,713 | 74,103 |
| WEST ELSDON | 9,488 | 76,870 | 75,555 |
| HYDE PARK | 35,884 | 68,959 | 75,748 |
| NORWOOD PARK | 50,328 | 122,019 | 78,520 |
| WEST TOWN | 95,409 | 122,647 | 80,506 |
| WEST LAWN | 8,552 | 83,756 | 81,180 |
| HEGEWISCH | 16,920 | 69,384 | 82,861 |
| NORTH CENTER | 151,814 | 179,675 | 83,298 |
| EAST SIDE | 3,577 | 68,331 | 88,385 |
| PULLMAN | 4,332 | 60,867 | 89,040 |
| AVALON PARK | 10,030 | 67,779 | 89,182 |
| UPTOWN | 34,882 | 93,765 | 89,974 |
| GREATER GRAND CROSSING | 2,445 | 34,362 | 92,791 |
| WEST GARFIELD PARK | 2,141 | 33,833 | 97,732 |
| NEAR NORTH SIDE | 162,924 | 162,683 | 99,394 |
| GAGE PARK | 5,240 | 62,139 | 100,210 |
| WEST RIDGE | 29,030 | 92,154 | 103,628 |
| KENWOOD | 35,979 | 93,229 | 104,763 |
| GRAND BOULEVARD | 16,095 | 50,588 | 106,412 |
| IRVING PARK | 49,450 | 133,718 | 106,704 |
| ALBANY PARK | 33,299 | 125,651 | 108,645 |
| NEAR WEST SIDE | 86,266 | 146,223 | 112,041 |
| JEFFERSON PARK | 57,108 | 175,395 | 113,056 |

*Table 4.3 (cont.)*

| Community | | | |
|---|---|---|---|
| AVONDALE | 41,927 | 135,022 | 124,993 |
| HUMBOLDT PARK | 8,025 | 85,886 | 126,287 |
| WEST ENGLEWOOD | 5,888 | 39,939 | 126,491 |
| BURNSIDE | / | 43,876 | 131,628 |
| NEAR SOUTH SIDE | 141,722 | 214,744 | 139,986 |
| NORTH LAWNDALE | 5,001 | 53,158 | 143,741 |
| GARFIELD RIDGE | 26,908 | 190,046 | 144,531 |
| DOUGLAS | 17,289 | 74,451 | 150,674 |
| RIVERDALE | 1,643 | 15,864 | 156,891 |
| LINCOLN SQUARE | 105,470 | 233,492 | 162,229 |
| ENGLEWOOD | 2,759 | 35,174 | 163,991 |
| FOREST GLEN | 237,864 | 277,241 | 176,208 |
| SOUTH DEERING | 10,478 | 71,752 | 182,718 |
| BRIGHTON PARK | 10,301 | 109,372 | 184,250 |
| BRIDGEPORT | 39,406 | 164,977 | 187,074 |
| NEW CITY | 12,321 | 135,178 | 216,485 |
| EAST GARFIELD PARK | 4,519 | 50,772 | 217,154 |
| AUSTIN | 24,355 | 64,247 | 240,966 |
| ARCHER HEIGHTS | 36,722 | 276,020 | 296,966 |
| ARMOUR SQUARE | 23,724 | 134,817 | 306,862 |
| MCKINLEY PARK | 45,833 | 301,481 | 416,580 |
| SOUTH LAWNDALE | 19,808 | 254,957 | 467,744 |
| LOWER WEST SIDE | 58,735 | 359,727 | 475,489 |
| NORTH PARK | 191,317 | 497,434 | 565,165 |
| FULLER PARK | / | 158,821 | 605,616 |
| CHICAGO LAWN | 20,796 | 206,121 | 655,970 |
| LAKE VIEW | 465,257 | 1,080,570 | 755,880 |
| SOUTH SHORE | 37,520 | 390,266 | 1,290,386 |
| LOOP | 5,212,926 | 4,155,762 | 1,948,920 |
| WASHINGTON PARK | 59,105 | 1,090,547 | 2,574,481 |
| WOODLAWN | 228,288 | 1,693,725 | 5,804,800 |
| SOUTH CHICAGO | 155,640 | 3,103,682 | 5,873,545 |
| DUNNING | 14,608,601 | 69,361,314 | 52,562,106 |
| OAKLAND | 31,288,384 | 847,158,896 | 1,964,383,414 |

**Table 4.4:** Accumulative accessibility score per capita for race groups by driving at the community level

| Community | Score of white population per capita | Score of black population per capita | Score of Hispanic population per capita | Score of Asian population per capita | Score of other races population per capita |
|---|---|---|---|---|---|

*Table 4.4 (cont.)*

| | | | | | |
|---|---|---|---|---|---|
| MOUNT GREENWOOD | 13,545 | 1,363 | 1,734 | 501 | 205 |
| OHARE | 9,787 | 228 | 2,290 | 985 | 411 |
| MONTCLARE | 4,333 | 796 | 7,381 | 492 | 298 |
| BEVERLY | 15,793 | 11,301 | 1,662 | 244 | 694 |
| WEST PULLMAN | 105 | 12,960 | 474 | 15 | 188 |
| MORGAN PARK | 6,053 | 13,518 | 1,083 | 25 | 539 |
| EDISON PARK | 34,298 | 886 | 3,602 | 1,287 | 842 |
| WASHINGTON HEIGHTS | 207 | 19,605 | 152 | 22 | 282 |
| ASHBURN | 2,746 | 14,395 | 9,240 | 190 | 649 |
| ROGERS PARK | 12,181 | 6,121 | 5,240 | 1,744 | 989 |
| CHATHAM | 410 | 28,917 | 285 | 247 | 252 |
| AUBURN GRESHAM | 178 | 23,578 | 478 | 55 | 185 |
| EDGEWATER | 28,166 | 5,850 | 7,187 | 5,221 | 1,883 |
| ROSELAND | 386 | 24,654 | 290 | 127 | 361 |
| PORTAGE PARK | 22,677 | 649 | 17,310 | 2,502 | 1,321 |
| CALUMET HEIGHTS | 513 | 35,827 | 1,461 | 47 | 195 |
| CLEARING | 15,420 | 855 | 24,031 | 177 | 671 |
| BELMONT CRAGIN | 3,288 | 1,229 | 25,815 | 431 | 181 |
| HERMOSA | 3,762 | 1,126 | 24,696 | 712 | 387 |
| LOGAN SQUARE | 55,813 | 4,110 | 24,211 | 4,707 | 3,181 |
| LINCOLN PARK | 116,500 | 7,088 | 9,548 | 11,004 | 4,995 |
| WEST ELSDON | 6,122 | 472 | 35,324 | 1,055 | 32 |
| HYDE PARK | 42,441 | 22,930 | 6,103 | 10,263 | 3,826 |
| NORWOOD PARK | 81,100 | 1,137 | 12,355 | 3,289 | 2,394 |
| WEST TOWN | 84,940 | 11,744 | 25,544 | 9,092 | 4,490 |
| WEST LAWN | 6,482 | 2,023 | 40,434 | 332 | 219 |
| HEGEWISCH | 25,114 | 2,923 | 35,251 | / | 93 |
| NORTH CENTER | 121,085 | 2,776 | 24,010 | 9,285 | 7,664 |
| EAST SIDE | 4,424 | 2,021 | 39,322 | 49 | 31 |
| PULLMAN | 2,959 | 64,158 | 1,773 | 19 | 782 |
| AVALON PARK | 418 | 64,116 | 361 | 490 | 1,162 |

*Table 4.4 (cont.)*

| | | | | | |
|---|---|---|---|---|---|
| UPTOWN | 65,863 | 17,015 | 12,841 | 11,922 | 3,739 |
| GREATER GRAND CROSSING | 471 | 49,846 | 691 | / | 655 |
| WEST GARFIELD PARK | 1,065 | 38,687 | 1,288 | 75 | 516 |
| NEAR NORTH SIDE | 176,937 | 14,160 | 15,132 | 45,421 | 8,078 |
| GAGE PARK | 1,692 | 2,149 | 37,690 | 318 | 30 |
| WEST RIDGE | 35,653 | 8,576 | 12,006 | 15,192 | 3,028 |
| KENWOOD | 21,414 | 81,111 | 2,949 | 14,134 | 5,250 |
| GRAND BOULEVARD | 2,970 | 73,648 | 1,386 | 604 | 1,392 |
| IRVING PARK | 47,986 | 2,987 | 44,889 | 9,912 | 4,203 |
| ALBANY PARK | 32,086 | 4,340 | 38,928 | 11,943 | 3,312 |
| NEAR WEST SIDE | 67,206 | 38,087 | 16,166 | 27,348 | 4,553 |
| JEFFERSON PARK | 84,596 | 1,557 | 30,003 | 11,488 | 3,142 |
| AVONDALE | 44,319 | 2,522 | 51,589 | 5,498 | 2,557 |
| HUMBOLDT PARK | 6,311 | 16,596 | 43,154 | 689 | 856 |
| WEST ENGLEWOOD | 682 | 52,331 | 4,430 | 10 | 740 |
| BURNSIDE | 336 | 76,197 | 1,245 | / | 807 |
| NEAR SOUTH SIDE | 131,673 | 67,055 | 12,170 | 60,948 | 9,664 |
| NORTH LAWNDALE | 2,641 | 49,486 | 12,409 | 106 | 487 |
| GARFIELD RIDGE | 29,422 | 16,162 | 68,890 | 5,518 | 1,356 |
| DOUGLAS | 12,808 | 66,746 | 4,094 | 21,499 | 3,632 |
| RIVERDALE | 1,411 | 58,690 | 1,308 | 137 | 4 |
| LINCOLN SQUARE | 152,977 | 7,372 | 30,148 | 17,767 | 10,029 |
| ENGLEWOOD | 733 | 72,136 | 4,223 | 266 | 323 |
| FOREST GLEN | 183,610 | 4,287 | 35,309 | 27,724 | 5,719 |
| SOUTH DEERING | 7,123 | 45,768 | 38,778 | 13 | 298 |
| BRIGHTON PARK | 5,280 | 1,141 | 68,382 | 8,572 | 839 |
| BRIDGEPORT | 48,845 | 4,291 | 36,044 | 58,693 | 2,000 |
| NEW CITY | 25,892 | 26,777 | 56,232 | 5,142 | 1,290 |
| EAST GARFIELD PARK | 6,068 | 81,464 | 3,257 | 495 | 1,172 |

*Table 4.4 (cont.)*

| | | | | | |
|---|---|---|---|---|---|
| AUSTIN | 3,827 | 101,745 | 6,781 | 148 | 985 |
| ARCHER HEIGHTS | 31,720 | 545 | 133,793 | 13,324 | 1,426 |
| ARMOUR SQUARE | 31,253 | 14,614 | 12,106 | 121,489 | 1,198 |
| MCKINLEY PARK | 37,619 | 2,596 | 130,477 | 75,707 | 1,774 |
| SOUTH LAWNDALE | 7,137 | 25,050 | 141,649 | 1,239 | 428 |
| LOWER WEST SIDE | 71,008 | 11,169 | 200,322 | 48,458 | 3,657 |
| NORTH PARK | 224,717 | 15,410 | 78,674 | 90,399 | 14,905 |
| FULLER PARK | 14,437 | 306,210 | 23,693 | 2,440 | 424 |
| CHICAGO LAWN | 6,127 | 207,838 | 48,173 | 1,379 | 15,822 |
| LAKE VIEW | 976,287 | 31,426 | 47,302 | 115,661 | 33,688 |
| SOUTH SHORE | 13,227 | 687,576 | 23,255 | 6,594 | 19,158 |
| LOOP | 3,798,473 | 677,393 | 438,997 | 981,636 | 248,595 |
| WASHINGTON PARK | 35,165 | 1,349,934 | 97,183 | 14,632 | 25,094 |
| WOODLAWN | 124,264 | 2,826,767 | 70,007 | 80,934 | 189,110 |
| SOUTH CHICAGO | 33,864 | 2,727,025 | 426,479 | 59,020 | 33,369 |
| DUNNING | 37,665,108 | 299,465 | 8,135,592 | 1,829,678 | 1,132,407 |
| OAKLAND | 4,014 | 1,297,002,058 | 4,781 | 2,356 | 11,377,774 |

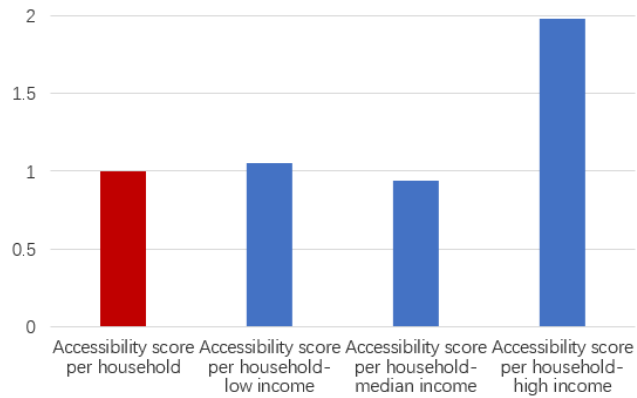

a. Accumulative accessibility score per household for income groups

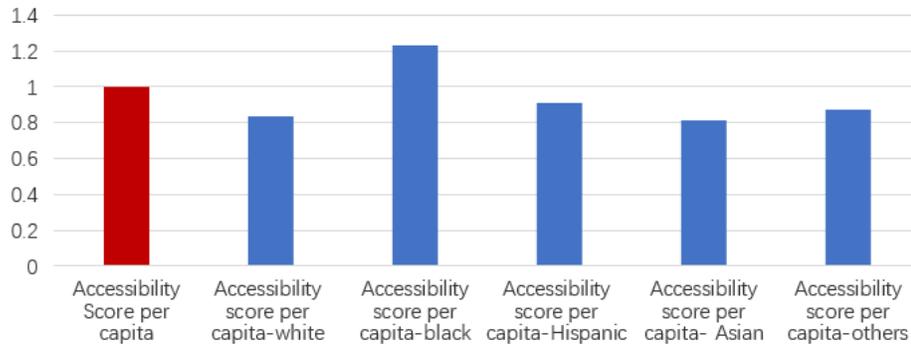

a. Accumulative accessibility score
per capita for race groups

**Figure 4.10:** Per capita cumulative accessibility scores to parks by driving of different population groups

Figure 4.7 a-c show the accessibility scores to parks of people in the low-, median-, and high-income groups, that were estimated using the gravity regression model. Within each land cell, the accessibility score and the different levels of income of the population affects the spatial distribution of the accessibility scores. The three maps correspond with these two factors. They coincide with the regression of the gravity model, in which the high scores tend to be distributed along Interstate 55 and 90. In addition, the difference in the accessibility score among the three levels of income corresponds with the distribution of the median level of income in Chicago as discussed above regarding the walking mode. Figure 4.7 d-h show the accessibility scores to parks of the gravity model for the different races. The spatial distribution of the accessibility scores for the races also corresponds to the factors that have been illustrated above regarding the walking mode.

Figure 4.8 explains the community level per capita accessibility score to parks by people driving. Several communities along the Chicago lakefront and in the west have high accessibility scores to parks, such as the Chicago Lawn and Lake View community.

Table 4.3 and 4.4 shows the community level per capita accessibility scores for different population groups. The accessibility scores for each group within each land cell is the same. However, when the land cells are combined and considered at the level of the community, the accessibility scores for every individual in a community will be different. Application of the gravity regression model resulted in the differences between different communities being vast. For the accessibility score of different races and income, the O'Hare community has the lowest accessibility to parks, and the Chicago Lawn Community has the best accessibility to parks. Within a community, the difference in accessibility of the races is generally larger than the difference in accessibility between the different income groups.

Figure 4.10 shows the accessibility score per household and per capita of different population groups. The high-income population also has the best access to parks by driving (Figure 4.10a). The Black population still has the highest average accessibility score to parks (Figure 4.10b).

**Comparison of Walking and Driving Modes**

A comparison of the maps of the accessibility scores to parks by people walking as estimated by the three different regression models (Figure 4.1 a-c), reveals that the areas with high accessibility scores are easy to find and are concentrated around the attractive parks. In contrast to this, when driving, the areas with high accessibility scores are likely to be concentrate along Interstates 90 and 55. and are convenient for access with transportation (Figure 4.6 a-c). People have limited access to parks by walking for 30 minutes, whereas people that are driving have much greater access to parks. Living close to an attractive park results in individuals having adequate access to parks on foot whereas living close to major roads will result in individuals that are driving having more access. A comparison of the average individual accessibility scores of the different groups by walking and driving (Figure 4.5 and Figure 4.10) reveals that the high-income and Black population have greater accessibility to parks by walking or driving than the other income or racial groups. In walking mode, the high-income population has an accessibility score that is only 1.3 times greater than the average access score (Figure 4.5a), whereas its accessibility score is nearly twice the average score in the driving mode (Figure 4.10a). The spatial distribution of the levels of income level in Chicago is illustrated in Figure 2.3. In the Central and North Sides of Chicago, many of the high-income populations are concentrate near Interstate 90, which means that they have convenient access to attractive parks by driving. However, in walking mode, the advantage of a transportation hub for the high-income population is not so obvious. Another difference is that when walking, the Hispanic population has a much lower accessibility score to parks than the other racial groups (Figure 4.5b), but it has relatively good accessibility when driving (Figure 4.10b). The reason can be explained with reference to the map of the distribution of the races in Chicago (Figure 2.4). Many White and Black populations live along Chicago's lakefront, which contains many highly attractive parks. In contrast to this, most of the Hispanic population live in the west of Chicago, and they do not have access to attractive parks along the lakefront by walking for 30 minutes. However, driving for 30 minutes a much greater distance can be covered than by walking for the same amount of time. Consequently, the access to attractive parks is not bad for Hispanic people who have a car.

# CHAPTER 5: ACCESSIBILITY EVALUATION PLANNING SUPPORTING SYSTEM (AEPSS)

In this part, a planning supporting system (PSS) is proposed based on the data we used in the present research. The accessibility evaluation planning supporting system (AEPSS) can not only be applied to evaluate the accessibility of parks, but can also be applied to evaluate all of the POIs in other cities. POIs include restaurants, shops, supermarkets, schools, hospitals, etc. If their rating and review data can be found on the Internet, such as Google Maps or Yelp, a similar analysis can be completed. With a planning supporting system, it is easy to change, interact, and manipulate data to get different results instead of repeating the same process every time. Accessibility is vital for measuring how reasonable the distribution of public facilities are. Cumulative accessibility of the population provides a better understanding of the attractiveness of services and the spatial distribution of facilities. The evaluation of the spatial distribution of cumulative accessibility can provide urban planners and policymakers with a basis for future planning and modification.

The AEPSS includes three core parts: a database, a model framework, and an interface that is used to visualize the analytical results and results in it being easy for users to interact with it. The database includes the road network, census data, landcover data, administrative boundary, data on the POIs (review number and rating, area). A summary of the analytical process of the model framework is as follows:

1.Input data:

Input data into the database (Figure 18).

2. Data analysis:

① Define the function to calculate the score for each POI (e.g. number of reviews, Review *Star of the rating).

② Choose a regression model to calculate the accumulated score for accessibility (e.g. gravity, kernel, and linear models).

③ Overlay the landcover data with the census data to obtain the distribution of the population in the research area.

④ Obtain the accumulated accessibility scores of each land cell.

⑤ Analyze the spatial accessibility for attractive POIs divided by area or by each land cell (Figure 19).

⑥ Compare the accessibility to POIs of different population groups (e.g. race, income)

⑦ Results:

● Regression map of the accumulated accessibility scores to attractive POIs (Figure 20).

● Cumulated accessibility score to POIs of different population groups (Figure 20).

● Diagram comparing the accumulated accessibility scores to POIs among different groups in the population (Figure 21).

3. Interface

The graphical user interface is used to visualize the analytical results and it is easy for users to interact with it.

The proposed interface to carry on the analysis as followed: (Figure 18- 21)

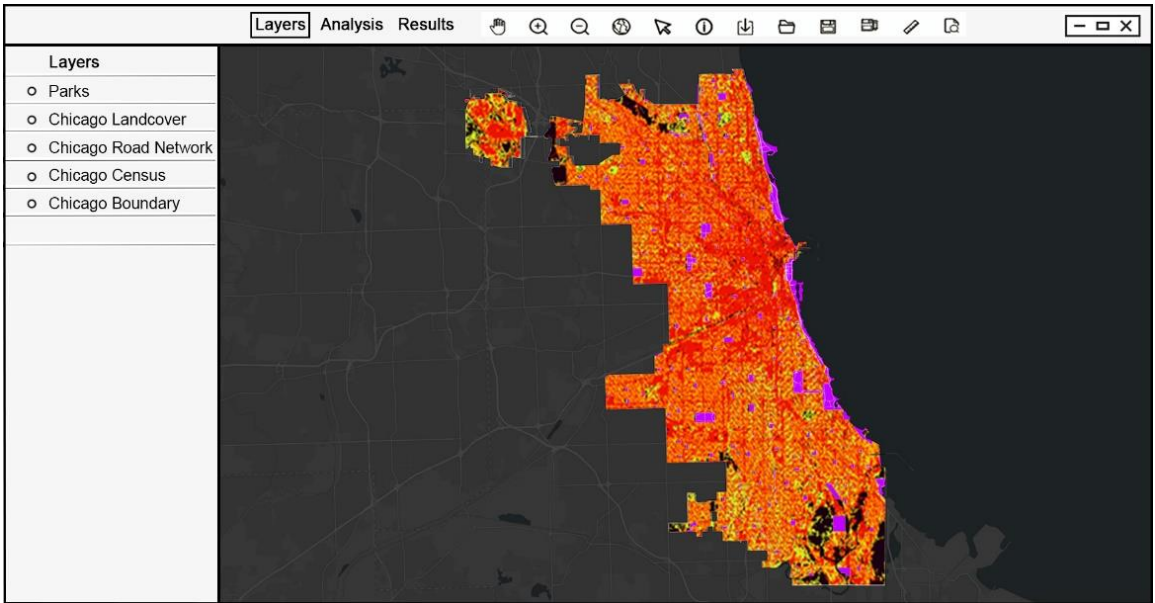
**Figure 5.1:** Input all required data layers

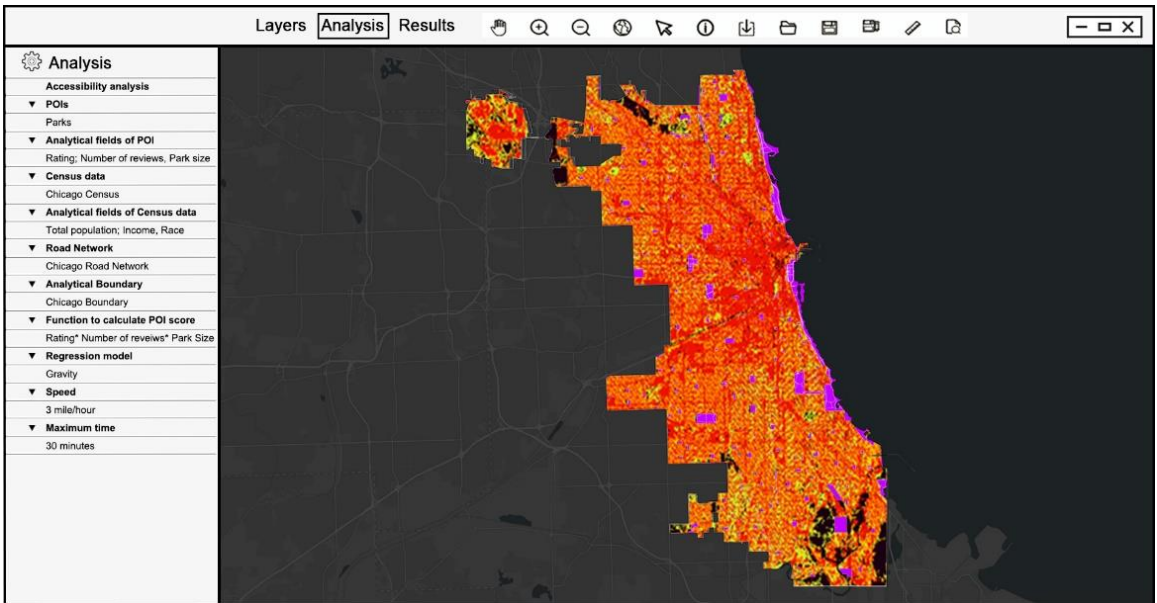
**Figure 5.2:** Set parameters and do the analysis

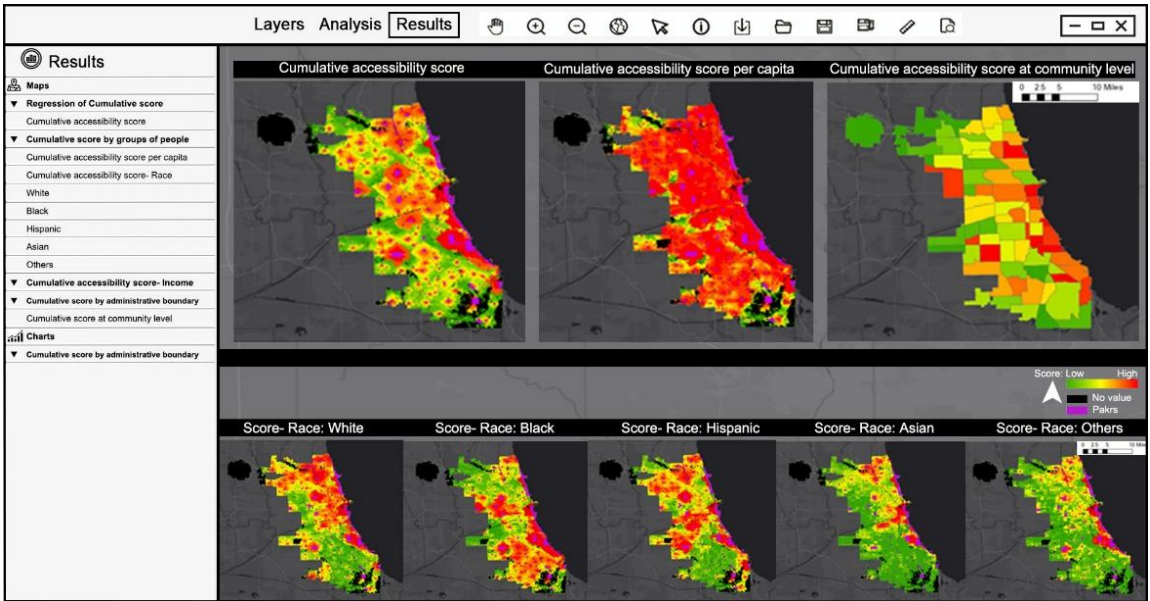
**Figure 5.3:** Results- Regression map of the accumulated accessibility scores to attractive POIs

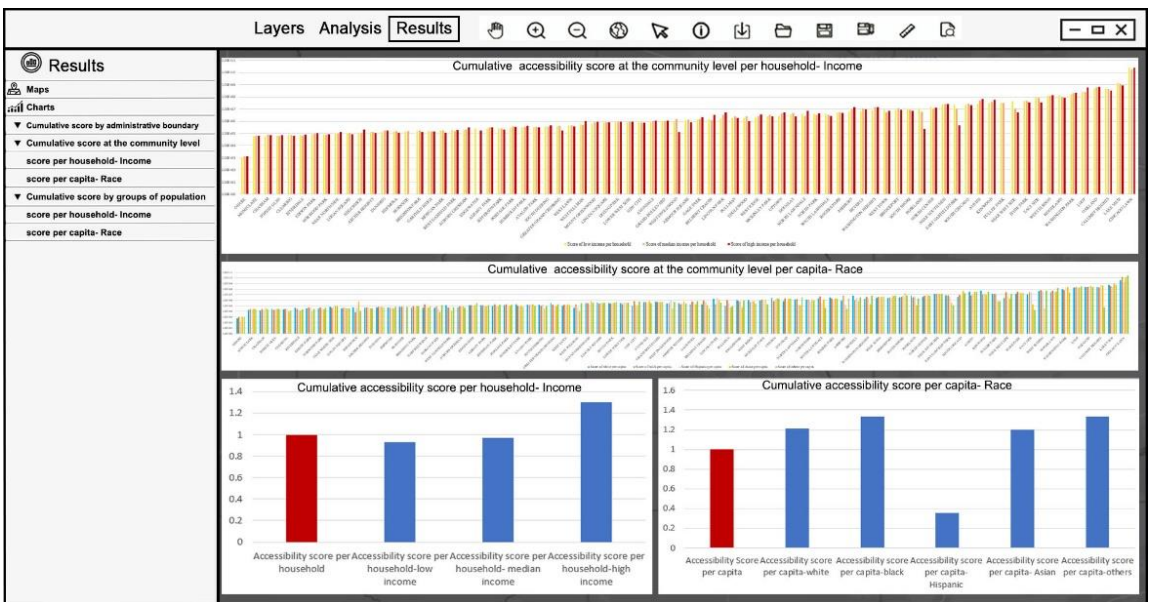
**Figure 5.4:** Results- Diagrams comparing the accumulated accessibility scores to POIs among different population groups

# CHAPTER 6: CONCLUSION

With the application of rating information from Google Maps, the present research aimed to evaluate the cumulative accessibility to parks of different population groups. To achieve this, the cumulative accessibility score was calculated for the different population groups. The results show that there are differences in people's ability to access parks based on the spatial distribution of the people and parks, their race, and their levels of income. Compared to traditional methods, the rating data were more convenient and efficient for evaluating people's preferences for the 583 parks in Chicago. Calculation of the accessibility scores for each 90 m X 90 m land cell provided detailed and accurate results and showed the regression tendencies more clearly. Moreover, the proposed planning supporting system can facilitate the process of evaluating all types of POIs in support of decision-making processes. For future research, the current AEPSS only provides the analysis and results of accessibility, so approaches for generating POIs selection need to be included in the AEPSS.

## Advantages, Limitations, and Future Research

The application of the rating and review data from Google Maps not only avoids the huge workload of requesting and collecting comments for individuals, but also reflects people's real altitudes towards each park, which makes it a more efficient and specific method. In terms of the limitation of network analysis and Euclidean distance in GIS, the application of the 2SFCA method to calculate the cumulative accessibility score reflects the detailed level of accessibility of each 90 m X 90 m land cell. In terms of the large size of Chicago, the map of the accessibility scores consists of continuous rather than categorical variables, and therefore it is better at accurately reflecting the actual situation and it is better for visualization. Based on the cumulative accessibility scores at the land cell level, the accessibility scores to parks at any scale larger than the land cell can be calculated, such as accessibility scores at the level of the community. The application of gravity, linear, and kernel regression models is useful, as they provide options for researchers to choose the method that is most suitable for their research. The planning supporting system is innovative because it can evaluate not only parks, but also all types of POIs for which rating information is available and can be efficiently gathered from the Internet.

There were limitations to the present research. When individuals are at the edge of a park, they should be treated as having reached the park. However, because in the present research the central point of each park was considered to be the park's location when calculating accessibility, individuals at the edge of a park were not yet considered to be in the park. This will cause some errors. Moreover, the results only provided an evaluation for current accessibility to parks and does not propose methods to evaluate the future quantitative selection of sites for public facilities. Consequently, it only reflects the current spatial distribution or assumed availability of POIs, and cannot determine the best locations for future developments.

International Conference on Multimedia Technology, 1376–1379.

https://doi.org/10.1109/ICMT.2011.6002845

Cord, A. f. ( 1 ), Schwarz, N. ( 1 ), & Roeßiger, F. ( 2 ). (2015a). Geocaching data as an indicator for recreational ecosystem services in urban areas: Exploring spatial gradients, preferences and motivations. Landscape and Urban Planning, 144, 151–162. https://doi.org/10.1016/j.landurbplan.2015.08.015

Cord, A. f. ( 1 ), Schwarz, N. ( 1 ), & Roeßiger, F. ( 2 ). (2015b). Geocaching data as an indicator for recreational ecosystem services in urban areas: Exploring spatial gradients, preferences and motivations. Landscape and Urban Planning, 144, 151–162. https://doi.org/10.1016/j.landurbplan.2015.08.015

Crilley, G. ( 1 ), Weber, D. ( 1 ), & Taplin, R. ( 2 ). (2012). Predicting visitor satisfaction in parks: Comparing the value of personal benefit attainment and service levels in Kakadu National park, Australia. Visitor Studies, 15(2), 217–237. https://doi.org/10.1080/10645578.2012.715038

Dadashpoor, H., Rostami, F., & Alizadeh, B. (2016). Is inequality in the distribution of urban facilities inequitable? Exploring a method for identifying spatial inequity in an Iranian city. Cities, 52, 159–172. https://doi.org/10.1016/j.cities.2015.12.007

Dai, D. (2011). Racial/ethnic and socioeconomic disparities in urban green space accessibility: Where to intervene? Landscape and Urban Planning, 102(4), 234–244. https://doi.org/10.1016/j.landurbplan.2011.05.002

Dai, D., & Wang, F. (2011). Geographic disparities in accessibility to food stores in southwest Mississippi. Environment and Planning B: Planning and Design, 38(4), 659–677. https://doi.org/10.1068/b36149

Dongsu KANG, Heechan KANG, Jeong Whon YU, Jinsu LEE, & Sunyoung PARK. (2015). An Integrated Model to Evaluate the Level-of-Service of Urban Rail Transfer Facilities in Consistent with User Perception. Journal of the Eastern Asia Society for Transportation Studies, 11, 1201. https://doi.org/10.11175/easts.11.1201

Dony, C. C., Delmelle, E. M., & Delmelle, E. C. (2015). Re-conceptualizing accessibility to parks in multi-modal cities: A Variable-width Floating Catchment Area (VFCA) method. Landscape and Urban Planning, 143, 90–99. https://doi.org/10.1016/j.landurbplan.2015.06.011

Egerer, M. ( 1, 2 ), Kendal, D. ( 3, 4 ), Ordóñez, C. ( 4 ), & Lin, B. b. ( 5 ). (2019). Multicultural gardeners and park users benefit from and attach diverse values to urban nature spaces. Urban Forestry and Urban Greening, 46. https://doi.org/10.1016/j.ufug.2019.126445

Farber, S., Bartholomew, K., Li, X., Páez, A., & Nurul Habib, K. M. (2014). Assessing social equity in distance based transit fares using a model of travel behavior. Transportation Research Part A: Policy and Practice, 67, 291–303. https://doi.org/10.1016/j.tra.2014.07.013

Gang, J., & Wolf, J. (2019). A New View, and a New Gateway, for Chicago. CTBUH Journal, 4, 12–19.

Hamilton, J. a. ( 1 ), Crompton, J. l. ( 1 ), & More, T. a. ( 2 ). (1991). Identifying the dimensions of service quality in a park context. Journal of Environmental Management, 32(3), 211–220. https://doi.org/10.1016/S0301-4797(05)80052-0

Hamstead, Z. A., Fisher, D., Ilieva, R. T., Wood, S. A., McPhearson, T., & Kremer, P. (2018). Geolocated social media as a rapid indicator of park visitation and equitable park access.